\documentclass[superscriptaddress,twocolumn,showpacs,a4paper,
amssymb,amsmath,nobibnotes,aps,prd,
showkeys,
nofootinbib,notitlepage]{revtex4-1}
\usepackage{verbatim}
\usepackage[T1]{fontenc}
\usepackage[utf8]{inputenc}
\usepackage[american]{babel}
\usepackage{epsfig}
\usepackage{graphicx,subcaption,caption}
\usepackage{booktabs}
\usepackage{multirow}
\usepackage{dcolumn}
\usepackage{amsmath}
\usepackage{mathtools}
\usepackage{amsfonts}
\usepackage{amssymb}
\usepackage{epstopdf}
\usepackage{bm}
\usepackage{siunitx}
\usepackage{braket}
\usepackage{enumitem}
\usepackage{soul}
\usepackage{color}

\ExplSyntaxOn 
\cs_new:Npn \expandableinput #1
  { \use:c { @@input } { \file_full_name:n {#1} } }
\ExplSyntaxOff

\makeatletter 
\providecommand*{\input@path}{}
\g@addto@macro\input@path{{"C:/Users/me/inpath/"}}
\makeatother

\definecolor{navyblue}{rgb}{0.0, 0.0, 0.5}
\definecolor{royalblue}{rgb}{0.25, 0.41, 0.88}
\definecolor{cadmiumgreen}{rgb}{0.0, 0.42, 0.24}
\definecolor{blue-violet}{rgb}{0.54, 0.17, 0.89}
\definecolor{darkviolet}{rgb}{0.58, 0.0, 0.83}
\definecolor{orange(colorwheel)}{rgb}{1.0, 0.5, 0.0}

\usepackage{hyperref}
\hypersetup{
    colorlinks=true, 
    linkcolor=royalblue, 
    citecolor=magenta}

\usepackage{booktabs}
\usepackage{multirow}
\usepackage{dcolumn}
\usepackage{colortbl}

\newcommand{\nq}[1]{%
	\begin{tabular}{@{}c@{}}\strut#1\strut\end{tabular}%
}

\begin{document}

\title{A novel model-marginalized cosmological bound on the QCD axion mass}

\date{\today}

\author{Eleonora Di Valentino}
\email{e.divalentino@sheffield.ac.uk}
\affiliation{School of Mathematics and Statistics, University of Sheffield, Hounsfield Road, Sheffield S3 7RH, United Kingdom}

\author{Stefano Gariazzo}
\email{gariazzo@to.infn.it}
\affiliation{Istituto Nazionale di Fisica Nucleare (INFN), Sezione di Torino, Via P.\ Giuria 1, I-10125 Turin, Italy}

\author{William Giar\`e}
\email{william.giare@uniroma1.it}
\affiliation{Galileo Galilei Institute for theoretical physics, Centro Nazionale INFN di Studi Avanzati, \\ Largo Enrico Fermi 2,  I-50125, Firenze, Italy}
\affiliation{INFN Sezione di Roma, P.le A. Moro 2, I-00185, Roma, Italy}

\author{Alessandro Melchiorri}
\email{alessandro.melchiorri@roma1.infn.it}
\affiliation{Physics Department and INFN, Universit\`a di Roma ``La Sapienza'', Ple Aldo Moro 2, 00185, Rome, Italy}

\author{Olga Mena}
\email{omena@ific.uv.es}
\affiliation{Instituto de F{\'\i}sica Corpuscular  (CSIC-Universitat de Val{\`e}ncia), E-46980 Paterna, Spain}

\author{Fabrizio Renzi}
\email{renzi@lorentz.leidenuniv.nl}
\affiliation{Lorentz Institute for Theoretical Physics, Leiden University, PO Box 9506, Leiden 2300 RA, The Netherlands}

\preprint{}

\begin{abstract}
We present model-marginalized limits on mixed hot dark matter scenarios, which consider both thermal neutrinos and thermal QCD axions. A novel aspect of our analyses is the inclusion of small-scale Cosmic Microwave Background (CMB) observations from  the Atacama Cosmology Telescope (ACT) and the South Pole Telescope (SPT), together with those from the Planck satellite and Baryon Acoustic Oscillation (BAO) data.  
After marginalizing over a number of well-motivated non-minimal background cosmologies, the tightest $95\%$~CL upper bound we obtain is $0.21$~eV, both for $\sum m_\nu$ and $m_{\rm a}$, from the combination of ACT, Planck and BAO measurements. Restricting the analyses to the standard $\Lambda$CDM picture, we find $\sum m_\nu<0.16$~eV and $m_{\rm a}<0.18$~eV, both at $95\%$~CL. Interestingly, the best background cosmology is never found within the minimal $\Lambda$CDM plus hot relics, regardless of the data sets exploited in the analyses. The combination of Planck with either BAO, SPT or ACT prefers a universe with a non-zero value of the running in the primordial power spectrum with strong evidence. Small-scale CMB probes, both alone and combined with BAO, either prefer,  with substantial evidence, non-flat universes (as in the case of SPT) or a model with a time varying dark energy component (as in the case of ACT). 
\end{abstract}


\maketitle

\section{Introduction}
\label{sec.Introduction}

The Peccei-Quinn (PQ) mechanism~\cite{Peccei:1977hh,Peccei:1977ur} represents the most elegant solution to the strong CP problem in Quantum Chromodynamics~\cite{Baker:2006ts,Pendlebury:2015lrz,Abel:2020pzs}. The key ingredient consists of a new dynamical pseudo-scalar field -- the \textit{axion}~\cite{Wilczek:1977pj,Weinberg:1977ma} -- which is driven towards its minimal energy configuration by the QCD dynamics, restoring the CP-invariance of strong interactions~\cite{Vafa:1984xg}.

The implications of a cosmic axion background crucially depend on the underlying production mechanism~\cite{DiLuzio:2020wdo}. If axions are produced via non-thermal channels, (\emph{e.g.} by the vacuum realignment mechanism~\cite{Abbott:1982af,Dine:1982ah,Preskill:1982cy,Linde:1985yf,Seckel:1985tj,Lyth:1989pb,Linde:1990yj} and/or by topological defects decay~\cite{Vilenkin:2000jqa,Kibble:1976sj,Kibble:1982dd,Vilenkin:1981kz,Davis:1986xc,Vilenkin:1982ks, Sikivie:1982qv,Vilenkin:1982ks,Huang:1985tt}) they should be considered natural cold dark matter candidates~\cite{Preskill:1982cy,Abbott:1982af,Dine:1982ah}. 
 
Conversely, a thermal population of axions produced by scattering and decays of particles can provide additional radiation energy-density contributing to the hot dark matter component of the Universe, similarly to massive neutrinos. Notice also that, while the axion cold dark matter density is a decreasing function of the mass, the axion couplings are proportional to the mass itself. In order to have a significant thermal population, axions must represent a sub-dominant component of the  total cold dark matter abundance and these two scenarios can be analyzed separately.

In this work we shall focus on the thermal axion mass limits from cosmology. A mandatory first step is the calculation of the axion relic abundance.

While most of the cosmological analyses carried out in the literature~\cite{Hannestad:2005df,Melchiorri:2007cd,Hannestad:2007dd,Hannestad:2008js,Hannestad:2010yi,Archidiacono:2013cha,Giusarma:2014zza,DiValentino:2015zta,DiValentino:2015wba,Archidiacono:2015mda,Ferreira:2020bpb} have been based on chiral perturbation theory, in Ref.~\cite{DiLuzio:2021vjd} it was pointed out that this approach can be safely extended only up to a temperature $T\lesssim 60$ MeV (see also Ref.~\cite{Giare:2020vzo,DiLuzio:2022tbb}), since the  perturbative scheme breaks down. A practical solution to settle this issue employs an interpolation of the thermalization rate to cover the gap between the highest safe temperature reachable by chiral perturbation theory and the regime above the confinement scale, where the axion production rate is instead dominated by the axion-gluon scattering~\cite{DEramo:2021psx,DEramo:2021lgb}. Improved cosmological bounds~\cite{DEramo:2022nvb} have been derived for two traditional benchmark models of QCD axion interactions namely, the KSVZ~\cite{Kim:1979if,Shifman:1979if} and the DFSZ one~\cite{Dine:1981rt,Zhitnitsky:1980tq} (see also Ref.~\cite{Caloni:2022uya}).\footnote{Although the KSVZ and DFSZ axion models are widely recognized as the most popular benchmark models~\cite{DiLuzio:2016sbl,DiLuzio:2017pfr}, it is worth considering numerous alternative models as well. Recent developments in this area can be found in Refs.\cite{Plakkot:2021xyx,Diehl:2023uui}. For a comprehensive review of these models, we refer to Ref.\cite{DiLuzio:2020wdo}.} As discussed in the same Ref.~\cite{DEramo:2022nvb}, the choice between KSVZ and DFSZ axion interaction scenarios does not result in a significant difference in the cosmological constraint on the axion mass, so that the current $95\%$~CL upper limits obtained in mixed hot dark matter scenarios in which massive neutrino species are also considered are $m_{\rm a}\lesssim 0.2$ eV and $\sum m_\nu\lesssim 0.15$~eV~\cite{DEramo:2022nvb}, using the most recent Cosmic Microwave Background (CMB) data released by the \textit{Planck} satellite~\cite{Aghanim:2019ame,Aghanim:2018eyx,Akrami:2018vks,Aghanim:2018oex}, the astrophysical observations of primordial light elements forged during the Big Bang Nucleosynthesis (BBN) epoch~\cite{Aver:2015iza,Peimbert:2016bdg,Cooke:2017cwo,Aghanim:2018eyx}, and the large scale structure information of the Universe in the form of Baryon Acoustic Oscillation (BAO) measurements~\cite{Beutler:2011hx,Ross:2014qpa,Alam:2016hwk}, see also the recent~\cite{Notari:2022zxo,DiLuzio:2022gsc}.

However, these limits have been obtained under two \emph{standardized} assumptions in cosmological parameter analyses. Namely, \textit{(1)} that the underlying model describing our universe is the minimal flat $\Lambda$CDM, and, \textit{(2)} restricting CMB measurements to those from the \textit{Planck} satellite observations. Concerning the first assumption, one should realize that a number of  several intriguing tensions and anomalies have emerged at different statistical levels~\cite{Abdalla:2022yfr,Perivolaropoulos:2021jda,DiValentino:2022fjm,DiValentino:2021izs}, questioning the validity of the canonical flat $\Lambda$CDM picture. A small curvature component, or a more general dark energy fluid, are some examples of very promising scenarios that should be carefully explored. From what regards the second assumption, analyses should also include the recent small-scale measurements of the CMB angular power spectra released by the Atacama Cosmology Telescope~\cite{ACT:2020frw,ACT:2020gnv} (ACT) and the South Pole Telescope~\cite{SPT-3G:2014dbx,SPT-3G:2021eoc} (SPT) Collaborations. 
It is therefore clear that the role of parameterizations, priors and models may lead to different constraints on the cosmological neutrino and axion masses.

Quantifying the impact resulting from the parameterization adopted for the cosmological model~\cite{diValentino:2022njd,Gariazzo:2018meg} and also from including independent CMB observations~\cite{DiValentino:2022oon,DiValentino:2022rdg} is the main goal of the present study, where, focusing exclusively on the KSVZ axion model, we derive new model-marginalized limits on hot dark matter scenarios involving axions $\&$ massive neutrinos. 
 
The paper is organized as follows: in \autoref{sec.Methods} we explain our statistical, computational and data analysis methods, in \autoref{sec.Results} we present the bounds on the hot dark matter masses in the different cosmological scenarios, together with the model-marginalized limits. We conclude in \autoref{sec.Conclusions}.

\section{Methodology}
\label{sec.Methods}

\subsection{Bayesian statistics}
The first aim of this study is to test how the results change when
an extended cosmological model is considered as the underlying theory,
instead of the simple $\Lambda$CDM scenario.
In order to do that, we proceed by performing a marginalization over a number of different models.\footnote{We would like to emphasize that our approach is to let the data determine whether a model is favored or disfavored, without introducing any pre-existing knowledge, based on both theoretical arguments or different observations, that could bias the results. This is to respect the principle that the theory should be informed by the data and not the other way around, and therefore we consider our approach to be both fair and conservative. }

Given a set of models ($\mathcal{M}_i$), in order to compute the model-marginalized posterior, one starts defining the posterior probabilities, $p_i$, of the model $\mathcal{M}_i$ over all the possible models:
\begin{equation}
\label{eq:modelposterior}
p_i = \frac{\pi_i Z_i}{\sum _j\pi_j Z_j}\,,
\end{equation}
where $\pi_i$ and $Z_i$ indicates the prior probability and the Bayesian evidence of model $\mathcal{M}_i$.
The model-marginalized posterior $p(\theta|d)$
for a set of parameters $\theta$, given some data $d$, reads as
\begin{equation}
\label{eq:mmposterior_def}
p(\theta|d)
\equiv
\sum_i
p(\theta|d,\mathcal{M}_i)
p_i
\,,
\end{equation}
where $p(\theta|d,\mathcal{M}_i)$ is the parameter posterior within the model $\mathcal{M}_i$. If all models have the same prior and using the Bayes factors $B_{i0}=Z_i/Z_0$ with respect to the favored model $\mathcal{M}_0$, the model-marginalized posterior is:
\begin{equation}
\label{eq:mmposterior}
p(\theta|d)
=
\frac{\sum_i
p(\theta|d,\mathcal{M}_i)
B_{i0}
}{
\sum_j B_{j0}
}
\,.
\end{equation}
Notice that if the Bayes factors are large in favor of the simplest and usually preferred model, extensions of the minimal picture will not contribute significantly to the model-marginalized posterior.
In order to perform Bayesian model comparison using the Bayes factors and evaluate the strength of preference in favor of the best model, we follow a modified version of the Jeffreys' scale\footnote{Notice that our empirical scale, summarized~\autoref{tab:bayes}, deviates from the scale defined in the original Jeffreys' work~\cite{Jeffreys:1939xee})} extracted from Ref.~\cite{Trotta:2008qt}, see~\autoref{tab:bayes}.

\begin{table*}[htbp!]
\begin{center}
\renewcommand{\arraystretch}{2}
\begin{tabular}{c | c  | c  | c  }
\hline
\textbf{$|\ln B_0|$} 
& \textbf{Odds} 
& \textbf{Probability} 
& \textbf{Strength of evidence}  
\\
\hline
$<0.1$& $\lesssim 3:1$ & $<0.750$ & Inconclusive\\
$1$& $\sim 3:1$ & $0.750$ & Weak\\
$2.5$& $\sim 12:1$ & $0.923$ & Moderate\\
$5$& $\sim 150:1$ & $0.993$ & Strong\\
\hline\hline
\end{tabular}
\end{center}
\caption{Modified Jeffreys’ empirical scale to establish the strength of evidence when comparing two competing models.}
\label{tab:bayes}
\end{table*}

At the time of determining neutrino mass bounds,
since the likelihood cannot put lower limits on the neutrino masses,
the prior range and shape can play a significant role,
as discussed for example in \cite{Heavens:2018adv,Stocker:2020nsx,Hergt:2021qlh,diValentino:2022njd}.
In order to avoid the dependency on prior in determining credible intervals, another possibility is to adopt the method of Ref.~\cite{Gariazzo:2019xhx}. Given some model $\mathcal{M}$ which contains a parameter $x$ (for instance, the axion or the neutrino mass),
the \emph{relative belief updating ratio} $\mathcal{R}(x_1, x_2|d,\mathcal{M})$ is defined as:
\begin{equation}
\label{eq:R_def}
\mathcal{R}(x_1, x_2|d,\mathcal{M})
\equiv
\frac{Z^{x_1}_\mathcal{M}}{Z^{x_2}_\mathcal{M}}
\,,
\end{equation}
where $Z^{x}_\mathcal{M}$
is defined as the Bayesian evidence of model $\mathcal{M}$ but fixing $x$ to a specific value:\footnote{We also assume that the prior on $x$ is independent on the other parameters and viceversa.}
\begin{equation}
\label{eq:Zx_def}
Z^{x}_\mathcal{M}
=
\int_{\Omega_\psi}
d\psi
\pi(\psi|\mathcal{M})
\mathcal{L}_\mathcal{M}(x,\psi)
\,,
\end{equation}
where $\psi$ represents all the parameters in model $\mathcal{M}$ except $x$,
which can vary in a parameter space $\Omega_\psi$,
$\pi(\psi|\mathcal{M})$ is their prior (notice that the $x$ prior is not included here)
and
$\mathcal{L}$ is the likelihood.

From Eq.~\eqref{eq:R_def}, we easily understand that the relative belief updating ratio
does not represent a probability,
as it is the ratio of two evidences. Importantly, the function $\mathcal{R}(x_1, x_2|d,\mathcal{M})$ is completely prior-independent.
Using the Bayes theorem,
it is possible to obtain a different expression for the former function:

\begin{equation}
\label{eq:R}
\mathcal{R}(x_1, x_2|d,\mathcal{M})
=
\frac{
p(x_1|d,\mathcal{M})/\pi(x_1|\mathcal{M})
}{
p(x_2|d,\mathcal{M})/\pi(x_2|\mathcal{M})
}\,,
\end{equation}
where $\pi(x|\mathcal{M})$ is the unidimensional prior on $x$.
This formulation is extremely useful in Monte Carlo Markov Chain (MCMC) runs where one can calculate these functions directly from the run chains. 
The definition of $\mathcal{R}(x_1,x_2|d)$
can be easily extended to perform a model marginalization:
\begin{equation}
\label{eq:mmR_def}
\mathcal{R}(x_1, x_2|d)
\equiv
\frac{
\sum_i Z^{x_1}_{\mathcal{M}_i}\pi(\mathcal{M}_i)
}{
\sum_j Z^{x_2}_{\mathcal{M}_j}\pi(\mathcal{M}_j)
}
\,,
\end{equation}
where now the evidences $Z^{x}_{\mathcal{M}_j}$ are computed within a specific model and
$\pi(\mathcal{M}_j)$ is the model prior.
In order to write $\mathcal{R}(x_1, x_2|d)$ using the parameter prior and posterior,
the simplest assumption is to consider the same prior $\pi(x)$ within all the models.
In such case, Eq.~\eqref{eq:mmR_def} becomes:
\begin{equation}
\label{eq:mmR}
\mathcal{R}(x_1, x_2|d)
=
\frac{
p(x_1|d)/\pi(x_1)
}{
p(x_2|d)/\pi(x_2)
}\,,
\end{equation}
where $p(x|d)$ is the model-marginalized posterior in Eq.~\eqref{eq:mmposterior_def}.

\subsection{Axion Modeling}
We address the effects induced by a relic population of thermal axions by employing a modified version of the Boltzmann integrator code \textsc{CAMB} \citep{Lewis:1999bs,Howlett:2012mh}. The code has been modified to accommodate the effects of QCD axions on cosmological scales only in terms of the axion mass which we employ as an additional cosmological parameters in our analysis, see Ref.~\cite{DEramo:2022nvb} for a detailed calculation.  We vary the axion mass in the range between $0.01$ eV and $10$ eV and focus exclusively on the KSVZ model of axion-hadron interactions. 

As long as the axion remains relativistic ($T_a\gg m_{\rm a}$), it behaves as radiation in the early Universe and its cosmological effects are those produced via their contribution to the effective number of neutrino species $N_{\rm eff}$. As detailed in Ref.~\cite{DEramo:2022nvb}, such a contribution is precisely evaluated by solving the Boltzmann equation for the axion number density in the early Universe. Indeed, very light axions ($m_{\rm a} \lesssim 0.1$ eV) are still relativistic at recombination and thus modify the CMB angular power spectrum via $N_{\rm eff}$. While such corrections are typically very small ($\Delta N_{\rm eff}\sim 0.03$), they are relevant for the next generation of CMB experiments~\cite{Abazajian:2016yjj}. On the other hand, heavier axions with masses larger than $0.1$ eV are highly non-relativistic at the recombination epoch. In this case, their impact on the CMB angular power spectra is both direct (through their impact on the early integrated Sachs-Wolfe effect, similarly to massive neutrinos) and indirect (by modifying the primordial helium abundance during the BBN). In this regard, it is worth stressing that the axion starts behaving as  cold dark matter much earlier than massive neutrinos, leading to a significant impact on structure formation. This feature, not only allows to distinguish massive neutrinos from massive axions through their effect on structure formation but it also allows to set stringent constraints on the axion mass exploiting large scale structure data, as well. Nonetheless, when the two species have similar masses, the evolution of their energy densities prevent to reach constraints on their masses lower than $ \sim 0.1$ eV, see also Refs.~\cite{Giare:2021cqr,DEramo:2022nvb}. Allowing for a prior on the axion mass spanning three orders of magnitude we can properly take into account all these effect of light and heavy axions on the CMB anisotropies, see ~\autoref{tab.Priors}.

\subsection{Cosmological Model Parameterization}

\begin{table}[htbp]
	\begin{center}
		\renewcommand{\arraystretch}{1.5}
		\begin{tabular}{l@{\hspace{0. cm}}@{\hspace{1.5 cm}} c}
			\hline
			\textbf{Parameter}    & \textbf{Prior} \\
			\hline\hline
			$\Omega_{\rm b} h^2$         & $[0.005\,,\,0.1]$ \\
			$\Omega_{\rm b} h^2$         & $[0.005\,,\,0.1]$ \\
			$\Omega_{\rm k} $     	     & $[-0.3\,,\,0.3]$\\
			$w_0$                        & $[-3\,,\,1]$ \\
			$w_a$                        & $[-3\,,\,2]$ \\
			$100\,\theta_{\rm {MC}}$     & $[0.5\,,\,10]$ \\
			$\log(10^{10}A_{\rm S})$     & $[2.91\,,\,3.91]$ \\
			$n_{\rm s}$                  & $[0.8\,,\, 1.2]$ \\
                $\alpha_{\rm s}$                  & $[-1\,,\, 1]$ \\
			$\sum m_{\nu}$ [eV]          & $[0.06\,,\,5]$\\
			$m_{\rm a}$ [eV]                   & $[0.01\,,\,10]$\\
			\hline\hline
		\end{tabular}
		\caption{List of uniform prior distributions for cosmological parameters.}
		\label{tab.Priors}
	\end{center}
\end{table}

As pointed out in the \hyperref[sec.Introduction]{Introduction}, a key point in our analysis is to derive robust bounds on the hot dark matter sector marginalizing over a plethora of possible background cosmologies. Therefore, along with the six $\Lambda$CDM parameters (the amplitude $A_s$ and the spectral index $n_s$ of scalar perturbations, the baryon $\Omega_b h^2$ and the cold dark matter $\Omega_c h^2$ energy densities, the angular size of the sound horizon at recombination $\theta_{\rm MC}$ and the reionization optical depth, $\tau$), we also include the sum of neutrino masses $\sum m_\nu$ and the axion mass $m_{\rm a}$. We then explore several extensions of this minimal model, enlarging the parameter space including one or more parameters, such as
a running of the scalar index ($\alpha_s$),
a curvature component ($\Omega_k$),
and
the dark energy equation of state parameters ($w_0$ and $w_a$)
(see \autoref{tab.Priors} for the priors adopted in the cosmological parameters). 
\begin{itemize}

    \item The running of scalar spectral index, $\alpha_s$. In simple inflationary models, the running of the spectral index is a second order perturbation and it is typically very small. However, specific models can produce a large running over a range of scale accessible to CMB experiments. Indeed, a non-zero value of $\alpha_s$ alleviates the $\sim 2.7\sigma$ discrepancy in the value of the scalar spectral index $n_{\rm s}$ measured by \emph{Planck} ($n_{\rm s}=0.9649\pm 0.0044$)~\cite{Planck:2018vyg,Forconi:2021que} and by the \emph{Atacama Cosmology Telescope}  (ACT) ($n_{\rm s}=1.008\pm 0.015$)~\cite{ACT:2020gnv}, see Refs.~\cite{DiValentino:2022rdg, DiValentino:2022oon,Giare:2022rvg}.

    \item Curvature density, $\Omega_k$. Recent data analyses of the CMB temperature and polarization spectra from Planck 2018 team exploiting the baseline \emph{Plik} likelihood suggest that our Universe could have a closed geometry at more than three standard deviations~\cite{Planck:2018vyg,Handley:2019tkm,DiValentino:2019qzk,Semenaite:2022unt}. These hints mostly arise from TT observations, that would otherwise show a lensing excess~\cite{DiValentino:2020hov,Calabrese:2008rt,DiValentino:2019dzu}. 
    In addition, analyses exploiting the \emph{CamSpec} TT likelihood~\cite{Efstathiou:2019mdh,Rosenberg:2022sdy} point to a closed geometry of the Universe with a significance above 99\% CL. Furthermore, an indication for a closed universe is also present in the BAO data, using Effective Field Theories of Large Scale Structure~\cite{Glanville:2022xes}. These recent findings strongly motivate to leave the curvature of the Universe as a free parameter~\cite{Anselmi:2022uvj} and obtain limits on the neutrino and axion masses in this context.

    \item Dynamical Dark Energy equation of state. 
    Cosmological neutrino and axion mass bounds become weaker if the dark energy equation of state is taken as a free parameter. Even if current data fits well with the assumption of a cosmological constant within the minimal $\Lambda$CDM scenario, the question of having an equation of state parameter different from $ -1 $ remains certainly open. Along with constant dark energy equation of state models, in this paper we also consider the possibility of having a time-varying $ w(a) $ described by the Chevalier-Polarski-Linder parametrizazion (CPL)~\cite{Chevallier:2000qy,Linder:2002et}:
    \begin{equation}
    \label{eq:cpl}
     	w(a) = w_0 + (1-a)w_a
    \end{equation}
    where $ a $ is the scale factor and is $ a_0 = 1 $ at the present time, $ w(a_0)=w_0 $ is the value of the equation of state parameter today. Dark energy changes the distance to the CMB consequently pushing it further (closer) if $w < -1$ ($w > -1$) from us. This effect can be balanced by having a larger matter density or, equivalently, by having more massive hot relics, leading to less stringent bounds on both the neutrino and axion masses. Accordingly, the mass bounds of cosmological neutrinos and axions become weaker if the dark energy equation of state is taken as a free parameter. 
\end{itemize}

\subsection{Statistical Analyses and Likelihoods}

In order to study the constraints achievable by current CMB and large scale structure probes, we make use of the publicly available code \textsc{COBAYA}~\citep{Torrado:2020xyz}. The code explores the posterior distributions of a given parameter space using the Monte Carlo Markov Chain (MCMC) sampler developed for \textsc{CosmoMC}~\cite{Lewis:2002ah} and tailored for parameter spaces with speed hierarchy implementing the ''fast dragging'' procedure developed in~\cite{Neal:2005}.  The prior distributions for the parameters involved in our analysis are chosen to be uniform along the range of variation (see \autoref{tab.Priors}) with the exception of the optical depth for which the prior distribution is chosen accordingly to the CMB datasets as discussed below. 
To perform model comparison, we compute the Bayesian Evidence of the different models and estimate the Bayes factors through the publicly available package \texttt{MCEvidence},\footnote{\href{https://github.com/yabebalFantaye/MCEvidence}{github.com/yabebalFantaye/MCEvidence}~\cite{Heavens:2017hkr,Heavens:2017afc}.} properly modified to be compatible with \textsc{COBAYA}.

We verified in some selected cases that the Bayes factors obtained with \texttt{MCEvidence}
are similar to those we obtained using \texttt{PolyChord}~\cite{Handley:2015fda,Handley:2015aa}, but much less time-consuming to obtain. We quantified the difference between \texttt{MCEvidence} and \texttt{PolyChord} results
by means of a set of dedicated simulations, for which we employed a 3D multi-modal Gaussian likelihood
to constrain a 3-parameter model as the simplest case, and compare it with two different models with 4 parameters.
The Bayesian evidences obtained with \texttt{MCEvidence} are systematically larger than those obtained with \texttt{PolyChord} by a factor of approximately $e$.
When computing the logarithm of the Bayes factors, the difference between \texttt{MCEvidence} and \texttt{PolyChord} ranges between -0.5 and 0.2 in the cases under consideration.
Therefore, we estimate that the values of the logarithms of the Bayes factors reported in the following have an uncertainty of 0.5 with respect to the values that one could have obtained with \texttt{PolyChord}. It is worth noting that the estimation of Bayes factors, starting from the MCMC, is weakly dependent on the chosen priors for cosmological parameters so that adopting uniform priors on $\sum m_{\nu}$ and $m_a$ may lead to differences in the resulting Bayesian Evidences. The impact of a uniform prior on $\sum m_{\nu}$ has been extensively discussed in literature, see e.g., Refs.~\cite{Simpson:2017qvj,Schwetz:2017fey,Gariazzo:2018pei,Gariazzo:2019xhx,Heavens:2018adv,GAMBITCosmologyWorkgroup:2020rmf,Hergt:2021qlh}. Concerning the axion mass, in our analysis, we focus on the cosmological thermal axion window and set a lower limit of $m_a\gtrsim 0.01$ eV in the prior. This range can be well explored using both linear and logarithmic sampling methods, and we checked that the choice of prior do not significantly affect the resulting limits or Bayesian Evidences.

Concerning the cosmological and astrophysical observations, our baseline data-sets and likelihoods include:

\begin{itemize}[leftmargin=*]

    \item Planck 2018 temperature and polarization (TT TE EE) likelihood, which also includes low multipole data ($\ell < 30$)~\citep{Aghanim:2019ame,Aghanim:2018eyx,Akrami:2018vks}. We refer to this combination as ``Planck 2018''.

    \item Planck 2018 temperature and polarization (TT TE EE) likelihood up to multipole $\ell=650$, to use in combination with the alternative ground-based small-scales CMB experiments. We refer to this combination as "Planck650".
	
	\item Planck 2018 lensing likelihood~\citep{Aghanim:2018oex}, reconstructed from measurements of the power spectrum of the lensing potential. We refer to this dataset as ``lensing''.
	
	\item Atacama Cosmology Telescope DR4 temperature and polarization (TT TE EE) likelihood, with a Gaussian prior on the optical depth at reionization $\tau = 0.065 \pm 0.015$, as done in~\cite{Aiola:2020azj}. We refer to this dataset combination as "ACT."
	
	\item South Pole Telescope polarization (TE EE) measurements SPT-3G~\cite{SPT-3G:2021eoc} combined with a Gaussian prior on the optical depth at reionization $\tau = 0.065 \pm 0.015$. We refer to this dataset combination as "SPT-3G."
	
	\item Baryon Acoustic Oscillations (BAO) measurements extracted from data from the 6dFGS~\cite{Beutler:2011hx}, SDSS MGS~\cite{Ross:2014qpa} and BOSS DR12~\cite{Alam:2016hwk} surveys. We refer to this dataset combination as ``BAO''.

\end{itemize}

\begin{figure*}
      \includegraphics[width =\textwidth]{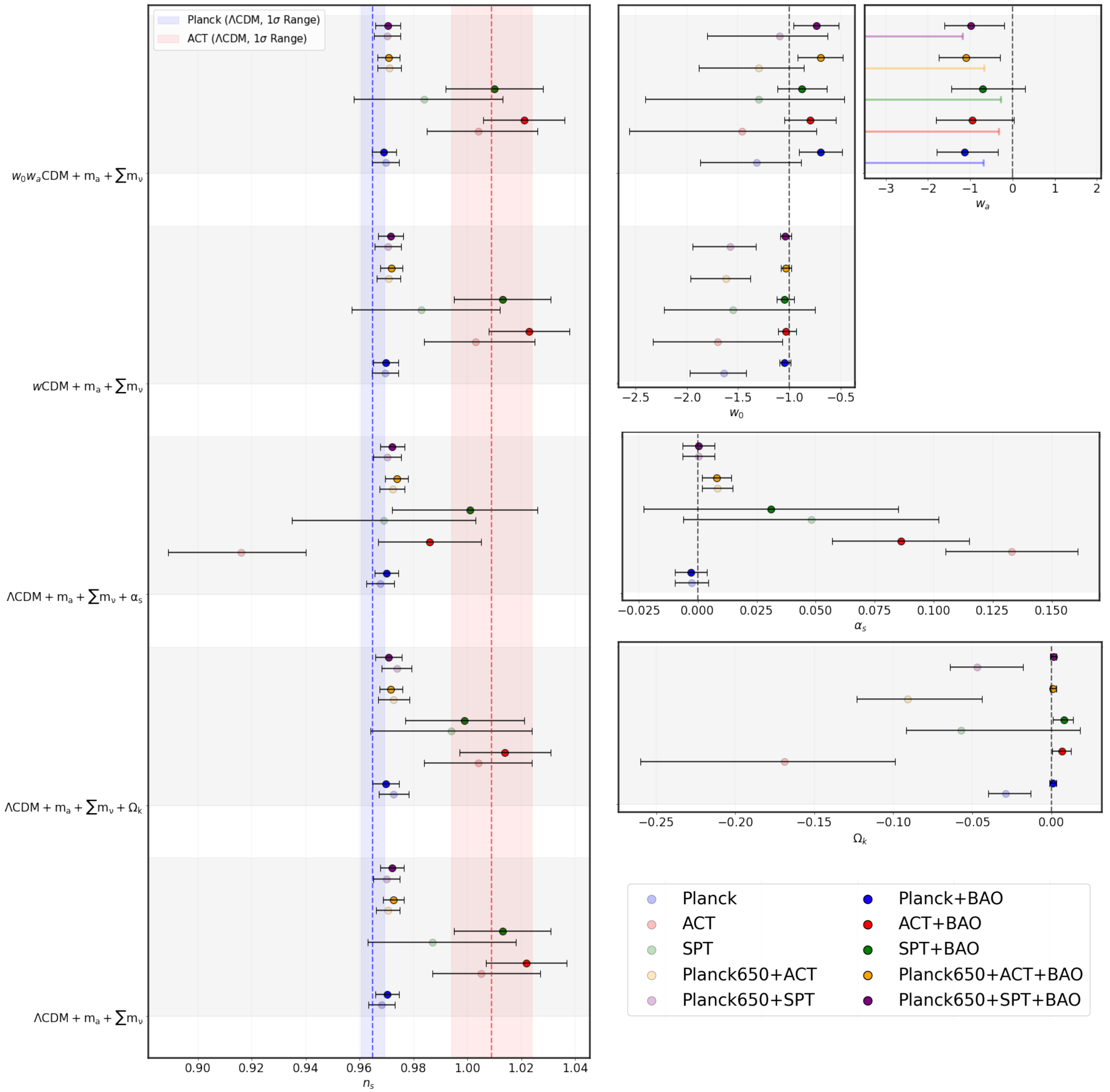} 
   \caption{Whisker plot with the mean values and their $68\%$~CL associated errors on $n_s$, $w_0$, $w_a$, $\alpha_s$ and $\Omega_k$ for different data combinations. The darker (lighter) circles depict the CMB limits with (without) the addition of BAO measurements. In the case of $n_s$ (left panel), we show the results for different background cosmologies, and the blue (red) vertical region refers to the value of $n_s$ as measured by Planck (ACT) within the baseline $\Lambda$CDM model.}
    \label{fig:whisker_all}
\end{figure*}

\begin{figure*}
\begin{tabular}{cc}
\multicolumn{2}{c}{ACT + Planck650 + BAO}\\
      \includegraphics[width = 0.5\textwidth]{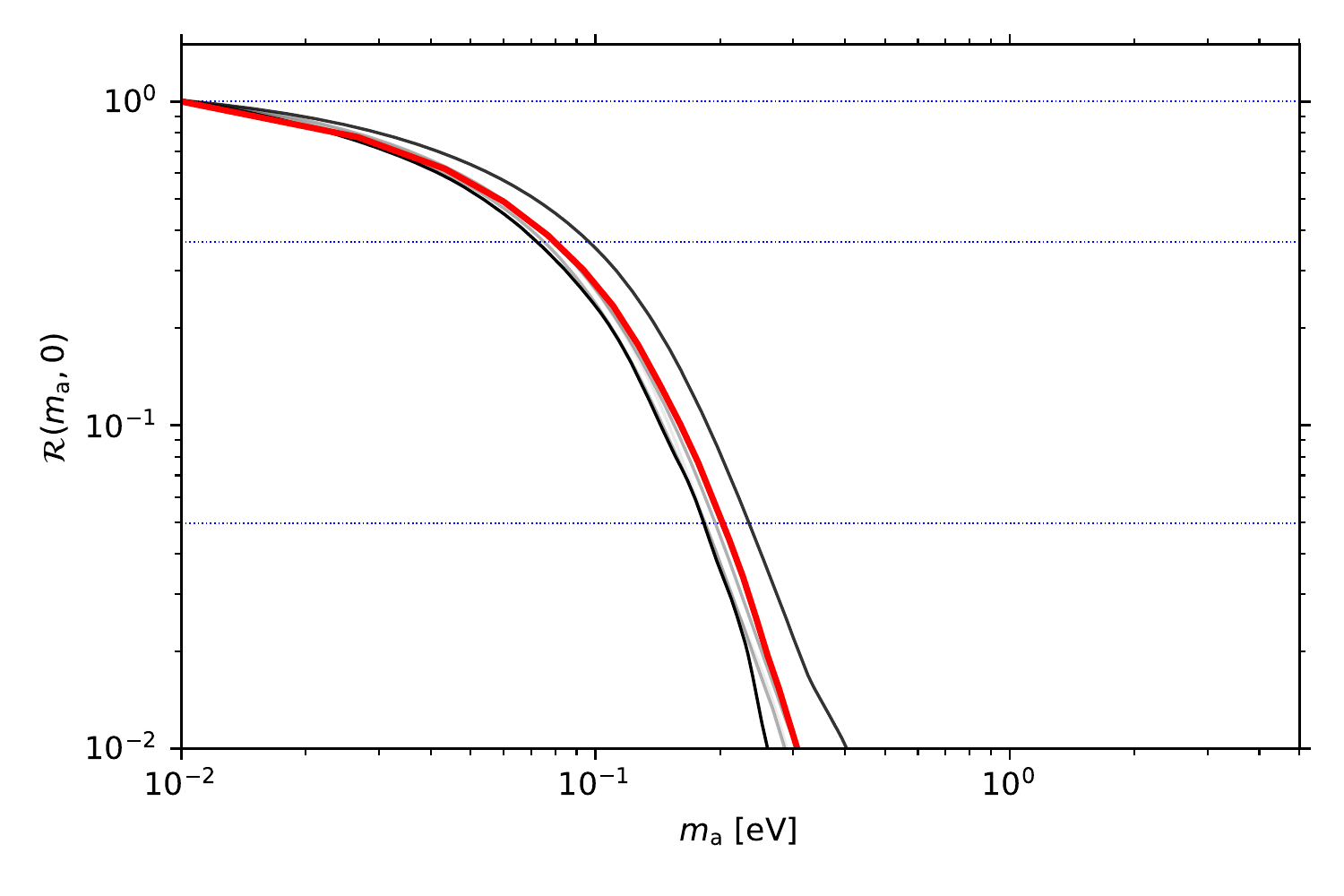} &
     \includegraphics[width = 0.5\textwidth]{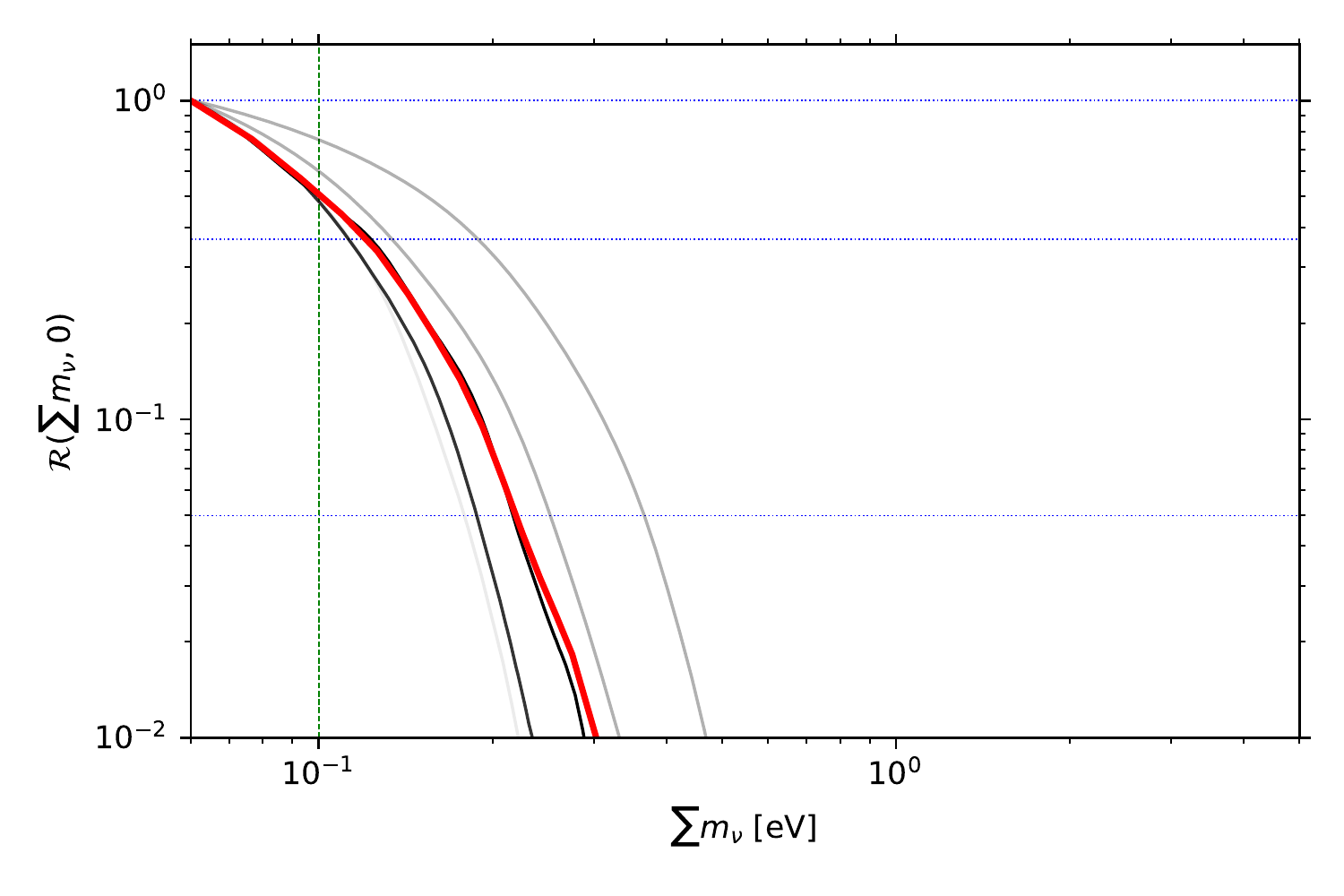}\\
\multicolumn{2}{c}{SPT + Planck650 + BAO}\\
     \includegraphics[width =
     0.5\textwidth]{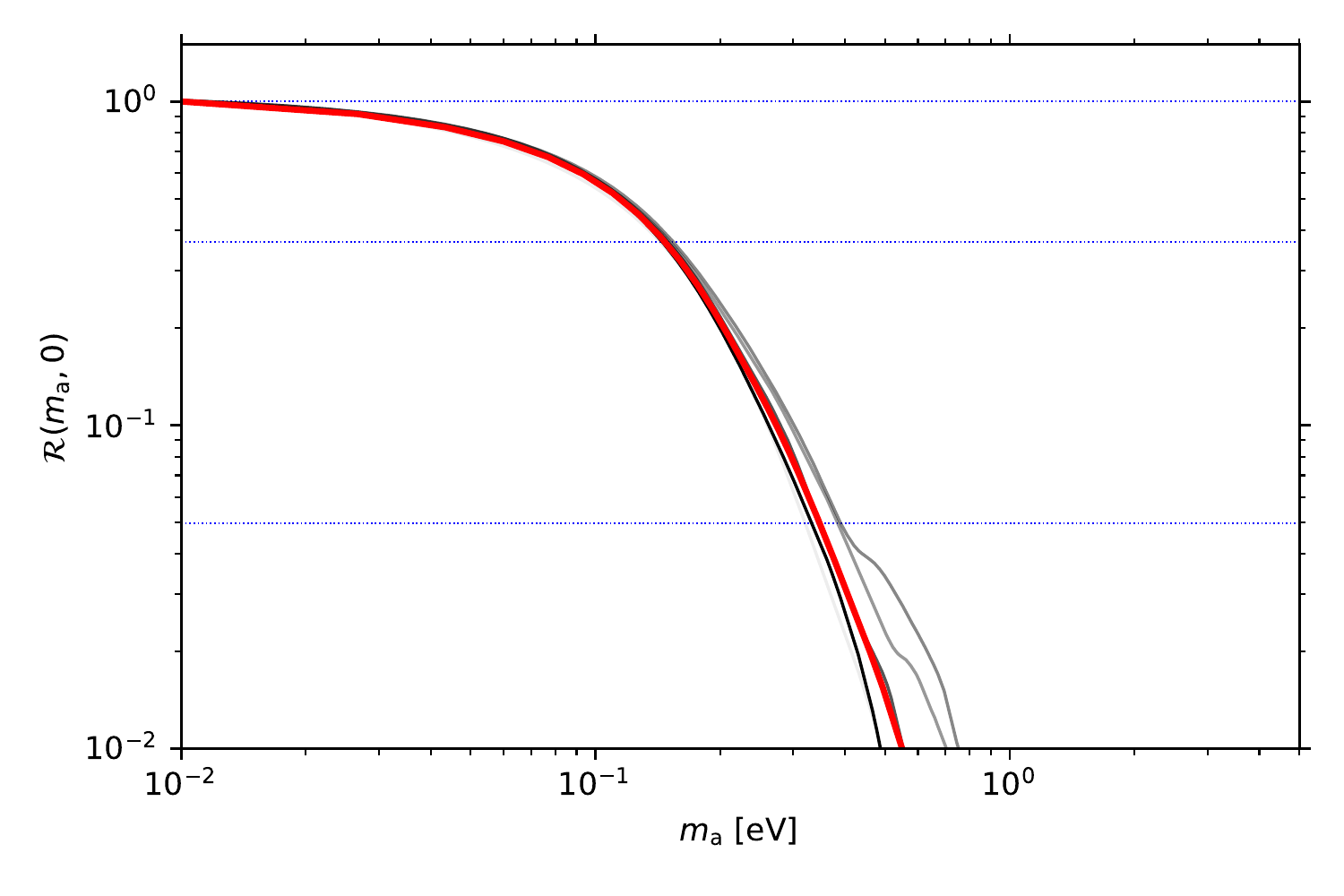} &
     \includegraphics[width = 0.5\textwidth]{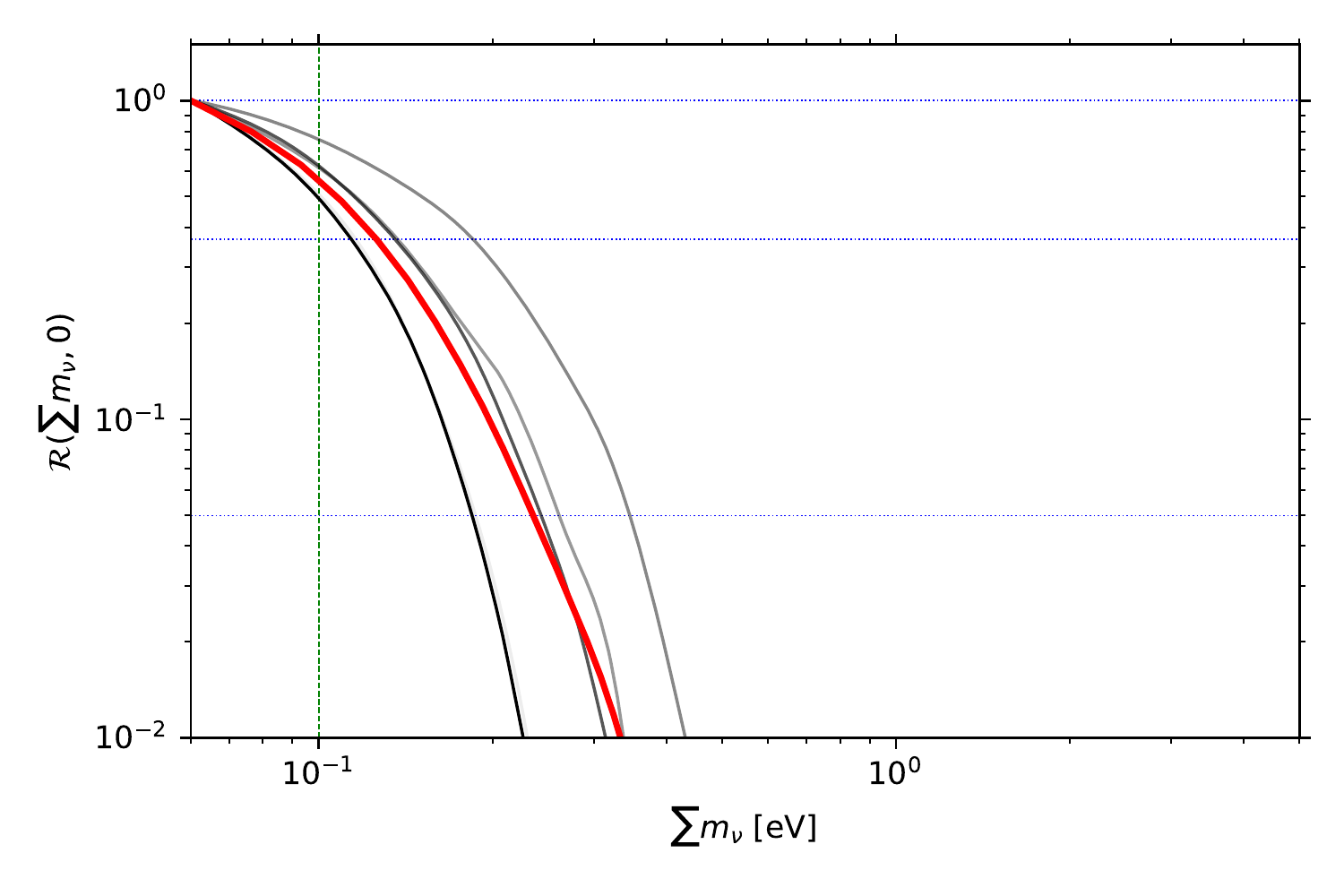}\\
      \end{tabular} 
   \caption{Model-marginalized relative belief updating ratio $\mathcal{R}$ for $m_{\rm a}$ (left) and $\sum m_\nu$ (right), considering the extensions of the $\Lambda$CDM model considered here. Black and gray lines show the $\mathcal{R}$ function within each model, where the darker lines are those that contribute most to the model marginalization, that is, they have the best Bayesian evidences. Horizontal lines show the significance levels $\exp(-1)$ and $\exp(-3)$. The upper (lower) panel refers to the ACT + Planck650 + BAO (SPT + Planck650 + BAO) data analyses. Vertical lines
indicate the value 0.1 eV, corresponding to the approximate lower limits for $\sum m_\nu$ in the inverted ordering case.}
    \label{fig:Rratio}
\end{figure*}

\begin{figure*}
      \includegraphics[width =\textwidth]{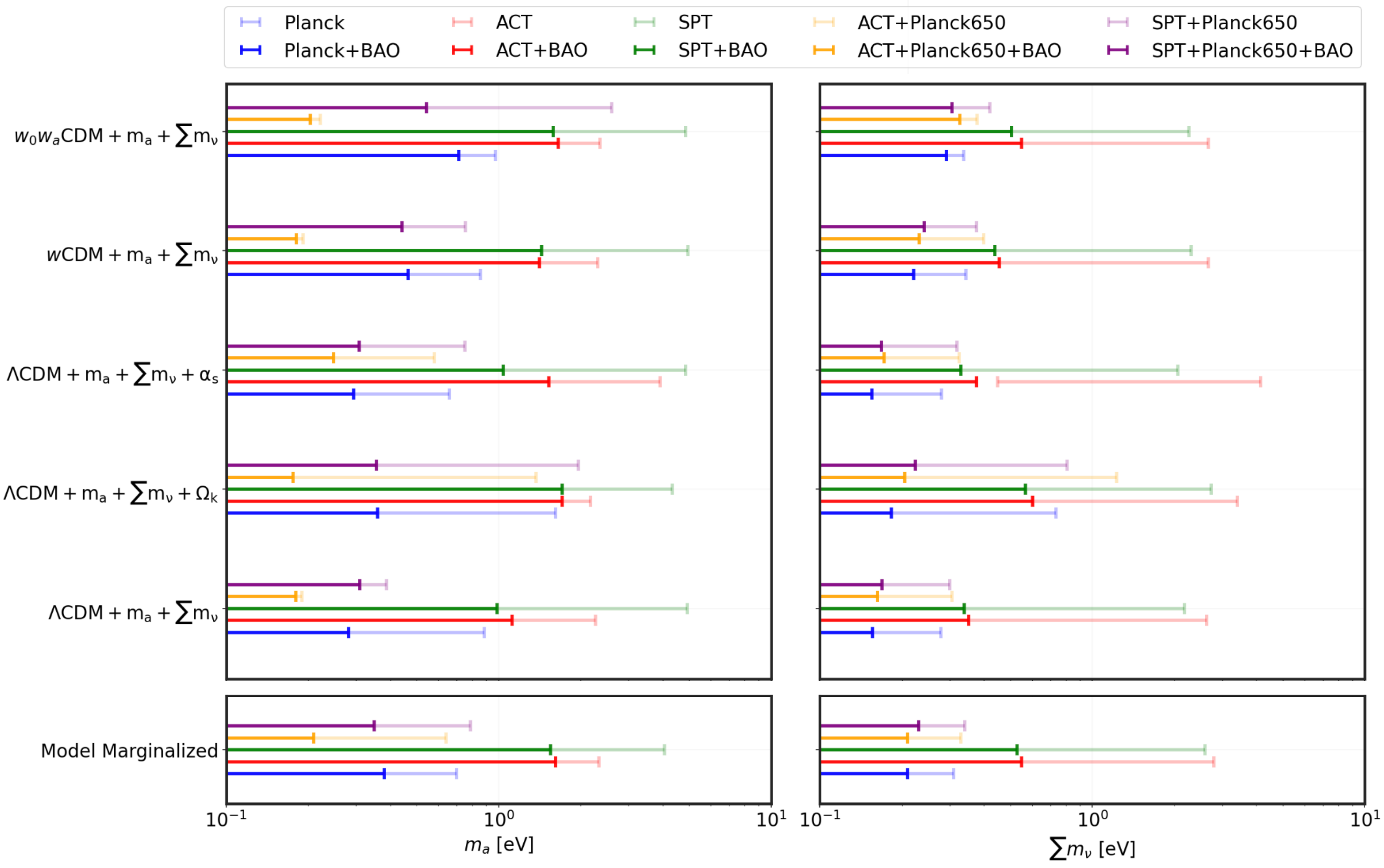} 
   \caption{Whisker plot with the $95\%$~CL upper bounds on the axion mass $m_{\rm a}$ (left) and on the total neutrino mass $\sum m_\nu$ (right) for different data combinations. The darker (lighter) lines depict the CMB limits with (without) the addition of BAO measurements. The top panels refer to constraints in each of the five possible background cosmologies explored here, while the lower panels show the model-marginalized ones derived here, see the main text of the manuscript for details.}
    \label{fig:whisker_ma_mnu}
\end{figure*}

\section{Results}
\label{sec.Results}
We start by discussing the limits in the mixed hot dark matter scenario assuming the standard $\Lambda$CDM cosmology. All the results for this case are provided in \hyperref[appendix.table]{Appendix A}, \autoref{tab.lCDM+mnu+ma}. The tightest constraints are obtained when combining Planck650 temperature, polarization and lensing measurements with ACT-CMB and BAO data: the limits we get on the hot dark matter relic masses are $m_{\rm a}<0.18$~eV and $m_\nu<0.16$~eV, both at $95\%$~CL. Adding ACT CMB observations therefore considerably improves the limit on hot relics, as with Planck plus BAO data alone the $95\%$~CL bounds are $m_{\rm a}<0.28$~eV and $m_\nu<0.16$~eV, in perfect agreement with the results for the KSVZ model of Ref.~\cite{DEramo:2022nvb} (see also the recent~\cite{Notari:2022zxo}). Concerning the remaining cosmological parameters, notice that both ACT and SPT observations (either alone or combined with BAO) prefer $n_s\simeq 1$, pointing to a Harrison-Zel'dovich primordial power spectrum, as can also be noticed from the left panel of  ~\autoref{fig:whisker_all} (see also Ref.~\cite{Giare:2022rvg}).

Enlarging the minimal $\Lambda$CDM picture with a curvature component $\Omega_k$ only degrades mildly the limits on $\sum m_\nu$, while the limit on $m_{\rm a}$ remains unchanged. 
From the results provided in \hyperref[appendix.table]{Appendix A}, \autoref{tab.omkCDM+mnu+ma} one can notice that the most constraining bounds are $m_{\rm a}<0.18$~eV and $m_\nu<0.20$~eV, both at $95\%$~CL for ACT plus Planck650 plus BAO observations. The preference for $n_s \simeq 1$ from SPT and ACT still persists, see the left panel of \autoref{fig:whisker_all}. Notice that \emph{all} CMB data prefer a value of $\Omega_k<0$ with a significance above the $\sim 2\sigma$ level for most of the cases. When CMB observations are combined with BAO measurements such a preference is however diluted. This behavior is shown in the bottom right panel of ~\autoref{fig:whisker_all}. 

Including a running ($\alpha_s$) of the scalar spectral index $n_s$, the $95\%$~CL bounds for the most powerful data set combination (i.e. Planck650 plus ACT and BAO) are $m_{\rm a}<0.25$~eV and $m_\nu<0.17$~eV, limiting the constraining power of these observations within the minimal $\Lambda$CDM scenario. Interestingly, the preference for $n_s\simeq 1$ from either SPT and/or ACT is not as strong as in the previous two background cosmologies (see the left panel of  ~\autoref{fig:whisker_all}) and it is instead translated into a mild preference for a non-zero value of $\alpha_s$ in the case of SPT. However, ACT observations shows a $\sim 5 \sigma $ preference for a positive value of $\alpha_s$, see the whisker plot in the right panel of ~\autoref{fig:whisker_all}, that corresponds to a preference for a positive neutrino mass. All the results for this case are provided in \hyperref[appendix.table]{Appendix A}, \autoref{tab.mnu+ma+nrun}.

We now leave freedom in the dark sector of the background cosmology. We start by discussing the simplest dark energy model with a constant dark energy equation of state $w_0$. The result for this case are provided in  \hyperref[appendix.table]{Appendix A}, \autoref{tab.w0}. First of all, notice that \emph{all} CMB measurements prefer a phantom dark energy universe, that is, a universe in which $w_0<-1$. The significance is larger than $2\sigma$ when considering Planck measurements, either alone or in combination with other CMB data sets. The larger negative value of $w_0$ is associated to a very large value of $H_0$, due to their strong degeneracy. Indeed, it has been shown that a phantom-like dark energy component can solve the current tension between high-redshift estimates and local universe measurements of the Hubble constant~\cite{DiValentino:2016hlg}. The addition of BAO observations leads however the value of $w_0$ very close to its cosmological constant expectation of $w_0=-1$ and the mean value of the Hubble constant is notably reduced, $H_0\sim 69$~km/s/Mpc. The results for $w_0$ are illustrated in the top right panel of ~\autoref{fig:whisker_all}. Concerning the limits on the hot relic masses, we obtain $m_{\rm a}<0.18$~eV for the axion mass and $\sum m_\nu <0.23$~eV for the neutrino masses, both at $95\%$~CL for the most powerful data set combination, which is, as in the previous background cosmologies, the one exploiting Planck650 plus ACT plus BAO observations. While the axion mass bound barely changes from the standard $\Lambda$CDM case, the neutrino mass limit is degraded to $\sum m_\nu <0.23$~eV, due to the strong degeneracy between the neutrino mass and the dark energy equation of state: if $w_0$ is allowed to freely change including also the phantom region, $\Omega_m$ can take very high values and also the neutrino mass can be much higher than in standard cosmological backgrounds. Our next step is to consider the widely exploited, two-parameter CPL parameterization for the dark energy component, see Eq.~(\ref{eq:cpl}). The results for this model are summarized in \hyperref[appendix.table]{Appendix A}, \autoref{tab.w0wa}. The constraints in for $w_0$ are very similar to those previously described, preferring all CMB observations values of $w_0<-1$ albeit with a mild significance. The corresponding $H_0$ value is also considerably larger than within the $\Lambda$CDM scenario (with hot relics). However, in this case, the addition of BAO data shifts the mean value of $w_0$ to the non-phantom region, with a very mild preference ($\sim 1.5\sigma$) for $w_0>-1$. Notice that CMB data alone is unable to measure the time derivative of the dark energy equation of state $w_a$, providing only an upper bound on this parameter. When BAO information is also considered in the analyses, a mean value of $w_a\sim -1$ is preferred.  The mean value of the Hubble constant after the inclusion of BAO observations is much closer to the value measured by the Planck collaboration in a standard cosmology. The results above for the dark energy parameters are illustrated by means of the whisker plots for the $w_0$ and $w_a$ parameters depicted in ~\autoref{fig:whisker_all}. 
Concerning the hot relics, notice that this background cosmology, having two extra parameters largely degenerated with the neutrino masses, leads to the least constraining hot relic mass bounds: the most powerful combination sets $95\%$~CL limits of $m_{\rm a}<0.20$~eV for the axion mass and $\sum m_\nu <0.33$~eV for the total neutrino mass.

\autoref{tab:BFs} presents the Bayes factors with respect to the best model for each of the five possible background cosmologies considered here and for the different data combinations. Interestingly, the best background cosmology is never found within the minimal $\Lambda$CDM plus two hot dark matter relics,\footnote{It is important to note that, for all datasets, $\Lambda$CDM is favored over the baseline hot relic extension that is considered in this work. This result is consistent with previous studies, such as Ref.\cite{diValentino:2022njd}, which discussed similar findings with regards to the effects of relic neutrinos.} regardless of the data set combinations. The combination of Planck or Planck650 with either BAO, SPT or ACT prefers a universe with a non-zero value of the running in the primordial power spectrum with strong evidence. Ground-based small-scale CMB probes, both alone and combined with BAO, prefer either non-flat universes, as in the case of SPT, or a model with a time varying dark energy component, as in the case of ACT. Such evidences are substantial when including BAO measurements.

\begin{table*}[htbp!]
\begin{center}
\renewcommand{\arraystretch}{2}
\resizebox{1 \textwidth}{!}{
\begin{tabular}{c | c c | c c | c c | c c | c c }
\hline
\textbf{Model} 
& \textbf{\nq{Planck}}  &  \textbf{\nq{\\ +BAO}} 
& \textbf{\nq{ACT}} & \textbf{\nq{\\ +BAO}}   
& \textbf{\nq{SPT}} & \textbf{\nq{\\ +BAO}}
& \textbf{\nq{ACT+Planck650}} & \textbf{\nq{\\ +BAO}} 
& \textbf{\nq{SPT+Planck650}} & \textbf{\nq{\\ +BAO}}
\\
\hline\hline

$\Lambda$CDM + $m_{\rm a}$ + $\sum m_\nu$&    6.73	&    6.46	&    0.25	&    3.06	&    1.43	&    3.38	&      4.71	&    5.06 &  4.41	&    5.29	\\

$\Lambda$CDM + $m_{\rm a}$ + $\sum m_\nu$ + $\alpha_s$&    0.00	&    0.00	&    6.35	&    4.39	&    
0.22	&    0.79	&  0.00	&    0.45	&   0.00	&    0.00	\\

$\Lambda$CDM + $m_{\rm a}$ + $\sum m_\nu$ +$\Omega_k$	&    5.15	&    0.13	&    1.84	&    0.80  & 0.00	&    0.00 &    7.38	&    0.00	&    5.51	&    0.82	\\

$w\Lambda$CDM + $m_{\rm a}$ + $\sum m_\nu$ 	&    5.50	&    1.64	&    0.26	&    0.78	&    2.75	&    1.66	&     5.37	&    2.37  &  5.55	&    1.82	\\

$w_0 w_a\Lambda$CDM + $m_{\rm a}$ + $\sum m_\nu$ 	&    5.35	&    1.62	&    0.00	&    0.00	&    2.36	&    0.70	&   5.68	&    2.37	&    7.29	&    1.51	\\

\hline \hline
\end{tabular}
}
\end{center}
\caption{Negative logarithms of the Bayes factors with respect to the best model for different data combinations, see also Appendix~\ref{sec:appendixBEBF}.}
\label{tab:BFs}
\end{table*}

\begin{table*}[htbp!]
\begin{center}
\renewcommand{\arraystretch}{2}
\begin{tabular}{c | c c |c c | c c| c c| c c}
\hline
\textbf{Parameter} 
& \textbf{\nq{Planck}}  &  \textbf{\nq{\\ +BAO}} 
& \textbf{\nq{ACT}} & \textbf{\nq{\\ +BAO}}   
& \textbf{\nq{SPT}} & \textbf{\nq{\\ +BAO}}
& \textbf{\nq{ACT+Planck650}} & \textbf{\nq{\\ +BAO}} 
& \textbf{\nq{SPT+Planck650}} & \textbf{\nq{\\ +BAO}}
\\
\hline\hline

$m_a$ [eV] (68 \%)	&    $<0.18$	&    $<0.14$	&    $<1.01$	&    $<0.71$	&    $<2.15$	&    $<0.69$	&      $<0.13$	&    $<0.09$ & $<0.20$	&    $<0.14$	\\

$m_a$ [eV] (95 \%)	&    $<0.70$	&    $<0.38$	&    $<2.33$	&    $<1.62$	&       $<4.06$	&    $<1.55$	&  $<0.64$	&    $<0.21$	& $<0.79$	&    $<0.35$	\\

$\sum m_\nu$ [eV] (68 \%)	&    $<0.16$	&    $<0.12$	&    $<1.42$	&    $<0.29$	&    $ <1.29$	&    $<0.29$	&   $<0.16$	&    $<0.12$	&   $<0.17$	&    $<0.13$	\\

$\sum m_\nu$ [eV] (95 \%)	&    $<0.31$	&    $<0.21$	&    $<2.79$	&    $<0.55$	&     $<2.59$	&    $<0.53$	& $<0.33$	&    $<0.21$	&     $<0.34$	&    $<0.23$	\\

\hline \hline
\end{tabular}
\end{center}
\caption{Marginalized upper bounds on $m_{\rm a}$ and $\sum m_{\nu}$ in eV for different data combinations.}
\label{tab:marglims}
\end{table*}

\autoref{fig:Rratio} shows the model-marginalized relative belief updating ratio $\mathcal{R}$, Eq.~(\ref{eq:mmR}), for both the axion mass $m_{\rm a}$ (left) and for the sum of the neutrino masses $\sum m_\nu$ (right), considering the extensions of the $\Lambda$CDM model considered and using ACT + Planck650 + BAO (SPT + Planck650 + BAO) data. The horizontal lines show the significance levels $\exp(-1)$ and $\exp(-3)$. 
The vertical lines 
indicate the value $0.1$~eV, corresponding to the approximate lower limits for $\sum m_\nu$ in the inverted ordering case.
The quantity $\mathcal{R}$ is independent of the shape and normalization of the prior and it is statistically equivalent to a Bayes factor between a model where $m_{\rm a}$ ($m_\nu$) has been fixed to some value and another model where $m_{\rm a}=0$ ($m_\nu=0$). The red curve shows the model-marginalized function $\mathcal{R}$, from which we derive the limits in~\autoref{tab:marglims}. The black and gray lines show the $\mathcal{R}$ function within each model, where the darker lines are those that contribute most to the model marginalization, that is, they have the best Bayesian evidences. For instance, for the case of ACT + Planck650+ BAO, the $95\%$~CL marginalized limit is $0.21$~eV, for both $m_{\rm a}$ and $\sum m_\nu$. Those bounds are led by the models which have the best Bayesian evidences, which, for this  particular data combination, are the 
$\Lambda$CDM + $m_{\rm a}$ + $\sum m_\nu$ + $\Omega_k$ and the 
$\Lambda$CDM + $m_{\rm a}$ + $\sum m_\nu$ + $\alpha_s$ ones, see ~\autoref{tab:BFs}, corresponding to the $95\%$~CL upper bounds of $m_{\rm a}<0.176$~eV, $\sum m_\nu <0.205$~eV and $m_{\rm a}<0.248$~eV, $\sum m_\nu <0.172$~eV, respectively.  
Instead, for the other data combination illustrated in ~\autoref{fig:Rratio}, that is,  SPT + Planck650 + BAO, 
the $95\%$~CL marginalized limits are $0.35$~eV and $0.23$, for $m_{\rm a}$ and $\sum m_\nu$ respectively. Those bounds are led by the models which have the best Bayesian evidences, which, for this  particular data combination,  are the 
$\Lambda$CDM + $m_{\rm a}$ + $\sum m_\nu$ + $\alpha_s$ and the 
$\Lambda$CDM + $m_{\rm a}$ + $\sum m_\nu$ + $\Omega_k$ ones, see ~\autoref{tab:BFs}, corresponding to the $95\%$~CL upper bounds of $m_{\rm a}<0.308$~eV, $\sum m_\nu <0.168$~eV and $m_{\rm a}<0.356$~eV, $\sum m_\nu <0.224$~eV, respectively. 
Interestingly, the minimal $\Lambda$CDM cosmology never provides the best Bayesian evidence, for any of these two data combinations. 
Notice also that, while the ACT + Planck650+ BAO data combination provides more powerful limits on $m_{\rm a}$ than the SPT + Planck650+ BAO one, these two data sets are equally powerful when constraining the neutrino mass, as can be noticed from the results shown in ~\autoref{tab:marglims}.

We conclude this section by summarizing our results in the whisker plots shown in ~\autoref{fig:whisker_ma_mnu}, illustrating   the $95\%$~CL upper bounds on the axion mass $m_{\rm a}$ and on the total neutrino mass $\sum m_\nu$ arising  for different data combinations in each of the five background cosmologies here. We also depict the model-marginalized limits on these two quantities. For the data combination Planck650 + ACT+ BAO, the most constraining bound for $m_{\rm a}$ is obtained within the $\Lambda$CDM + $m_{\rm a}$ + $\sum m_\nu$ + $\Omega_k$ scenario ($m_{\rm a} < 0.176$~eV at $95\%$~CL). For the total neutrino mass, the tightest $95\%$~CL upper bound ($m_{\rm a} < 0.163$~eV at $95\%$~CL) is found in the $\Lambda$CDM + $m_{\rm a}$ + $\sum m_\nu$ canonical scheme. For the data set  SPT + Planck650+ BAO, the tightest limits on the hot thermal relic masses are those derived in the $\Lambda$CDM + $m_{\rm a}$ + $\sum m_\nu$ +
$\alpha_s$ cosmological background, and correspond to $m_{\rm a} < 0.301$~eV and $\sum m_\nu < 0.168$~eV (both at $95\%$~CL).

\section{Conclusions}
\label{sec.Conclusions}
Axions provide the most elegant solution to the strong CP problem in Quantum Chromodynamics. In the early universe, axions can be produced via thermal or non thermal processes. Indeed, an axion population produced by scattering and decays of particles can provide additional radiation energy-density contributing to the hot dark matter component of the Universe, similarly to massive neutrinos. Therefore, it is certainly possible to set thermal axion mass limits from cosmology.
Previous works in the literature have computed the current thermal axion population based on chiral perturbation theory. However, these limits can not be extended to high temperatures in the early universe because the underlying perturbation theory would not longer be valid. A possible method to overcome this problem makes use of an interpolation of the thermalization rate in order to cover the gap between the highest safe temperature reachable by chiral perturbation theory and the regime above the confinement scale, where the axion production rate is instead dominated by the axion-gluon scattering~\cite{DEramo:2021psx,DEramo:2021lgb}. 

Nevertheless, all previous axion mass bounds in the literature assume the minimal flat $\Lambda$CDM and neglect the other ground-based small-scale CMB measurements than those of \textit{Planck} satellite observations. 

Here we relax the two above assumptions and present novel model-marginalized limits on mixed hot dark matter scenarios, which consider both thermal neutrinos and thermal QCD axions. A new aspect of our analyses is the inclusion of small-scale Cosmic Microwave Background (CMB) observations from  the Atacama Cosmology Telescope (ACT) and the South Pole Telescope (SPT), together with those from the Planck satellite and Baryon Acoustic Oscillation (BAO) data.  
The tightest $95\%$~CL marginalised limits are $0.21$~eV, for both $\sum m_\nu$ and $m_{\rm a}$, from the combination of ACT, Planck650 and BAO measurements. Restricting the analyses to the standard $\Lambda$CDM picture extended with free neutrino and axion masses, we find $\sum m_\nu<0.16$~eV and $m_{\rm a}<0.18$~eV, both at $95\%$~CL. Interestingly, the best background cosmology is never found within the minimal $\Lambda$CDM plus hot relics, regardless of the data sets exploited in the analyses. The combination of Planck or Planck 650 with either BAO, SPT or ACT prefers a universe with a non-zero value of the running in the primordial power spectrum with strong evidence. Ground-based small-scale CMB probes, both alone and combined with BAO, prefer either with substantial evidence for non-flat universes, as in the case of SPT, or a model with a time varying dark energy component, as in the case of ACT.
If the existence of an axion which may be thermally produced in the early universe and neutrino masses will be independently confirmed by other probes, upcoming cosmological observations may strengthen the evidence against the minimal cosmological framework, pointing to possible exciting new ingredients in the theory.

\begin{acknowledgments}
\noindent  
EDV is supported by a Royal Society Dorothy Hodgkin Research Fellowship. 
This article is based upon work from COST Action CA21136 Addressing observational tensions in cosmology with systematics and fundamental physics (CosmoVerse) supported by COST (European Cooperation in Science and Technology). AM and WG are supported by the TASP INFN initiative.
We acknowledge IT Services at The University of Sheffield for the provision of services for High Performance Computing.
This work has been partially supported by the MCIN/AEI/10.13039/501100011033 of Spain under grant PID2020-113644GB-I00, by the Generalitat Valenciana of Spain under grant PROMETEO/2019/083 and by the European Union’s Framework Programme for Research and Innovation Horizon 2020 (2014–2020) under grant agreement 754496 (FELLINI) and 860881 (HIDDeN).
\end{acknowledgments}

\section*{Data availability}
The data and chains underlying this article will be shared on reasonable request to the corresponding author.

\appendix
\section{Tables}

In this appendix we provide the tables with all the results for the cosmological parameters for all the models discussed in our work:


\begin{center}
\renewcommand{\arraystretch}{1.8}
\begin{tabular}{l@{\hspace{2.5 cm}} c}
\toprule
\textbf{Cosmological Model}    & \textbf{Results in} \\
$\Lambda$CDM+$\sum m_{\nu}$+$m_a$                  & \autoref{tab.lCDM+mnu+ma} \\
$\Lambda$CDM+$\sum m_{\nu}$+$m_a$+$\Omega_k$       & \autoref{tab.omkCDM+mnu+ma} \\
$\Lambda$CDM+$\sum m_{\nu}$+$m_a$+$\alpha_s$       & \autoref{tab.mnu+ma+nrun}\\
$w$CDM+$\sum m_{\nu}$+$m_a$                        & \autoref{tab.w0} \\
$w_0w_a$CDM+$\sum m_{\nu}$+$m_a$                   & \autoref{tab.w0wa} \\

\bottomrule
\end{tabular}
\end{center}

\label{appendix.table}
\begin{table*}[htbp!]
	\begin{center}
		\renewcommand{\arraystretch}{2}
		\resizebox{1 \textwidth}{!}{\begin{tabular}{c | c c| c c| c  c|  c c| c c  }

				\hline
				\textbf{Parameter} & \textbf{\nq{Planck}}  &  \textbf{\nq{\\ +BAO}} & \textbf{\nq{ACT}} & \textbf{\nq{\\ +BAO}}  & \textbf{\nq{SPT}} & \textbf{\nq{\\ +BAO}}  & \textbf{ACT+Planck650} & \nq{\\ \textbf{+BAO}} & \textbf{SPT+Planck650} & \nq{\\ \textbf{+ BAO}} \\
				\hline\hline

$ \Omega_\mathrm{b} h^2  $ & $  0.02243\pm 0.00015 $ & $  0.02250\pm 0.00015 $ & $  0.02165\pm 0.00033 $ & $  0.02170\pm 0.00032 $ & $  0.02251\pm 0.00033 $ & $  0.02252\pm 0.00032 $ & $  0.02238\pm 0.00014 $ & $  0.02245\pm 0.00013 $ & $  0.02246\pm 0.00014 $ & $  0.02252\pm 0.00013 $ \\ 
$ \Omega_\mathrm{c} h^2  $ & $  0.1225^{+0.0016}_{-0.0022} $ & $  0.1208^{+0.0012}_{-0.0014} $ & $  0.1242\pm 0.0046 $ & $  0.1190^{+0.0024}_{-0.0020} $ & $  0.1195^{+0.0060}_{-0.0046} $ & $  0.1175^{+0.0022}_{-0.0020} $ & $  0.1214\pm 0.0015 $ & $  0.1204^{+0.0011}_{-0.0012} $ & $  0.1216^{+0.0014}_{-0.0018} $ & $  0.1204^{+0.0012}_{-0.0014} $ \\ 
$ \tau  $ & $  0.0564\pm 0.0076 $ & $  0.0581^{+0.0070}_{-0.0080} $ & $  0.072\pm 0.015 $ & $  0.070\pm 0.015 $ & $  0.067\pm 0.015 $ & $  0.069\pm 0.014 $ & $  0.0562^{+0.0072}_{-0.0082} $ & $  0.0567^{+0.0070}_{-0.0081} $ & $  0.0548\pm 0.0081 $ & $  0.0556\pm 0.0078 $ \\ 
$ 100\theta_\mathrm{MC}  $ & $  1.04058\pm 0.00036 $ & $  1.04081\pm 0.00032 $ & $  1.04120\pm 0.00081 $ & $  1.04208\pm 0.00064 $ & $  1.03854\pm 0.00082 $ & $  1.03939\pm 0.00066 $ & $  1.04087\pm 0.00031 $ & $  1.04101\pm 0.00027 $ & $  1.04044\pm 0.00032 $ & $  1.04060\pm 0.00029 $ \\ 
$ n_\mathrm{s}  $ & $  0.9681\pm 0.0049 $ & $  0.9703\pm 0.0043 $ & $  1.005^{+0.022}_{-0.018} $ & $  1.022\pm 0.015 $ & $  0.987^{+0.031}_{-0.024} $ & $  1.013\pm 0.018 $ & $  0.9705\pm 0.0043 $ & $  0.9727^{+0.0037}_{-0.0041} $ & $  0.9700\pm 0.0048 $ & $  0.9721\pm 0.0043 $ \\ 
$ \log(10^{10} A_\mathrm{s})  $ & $  3.055^{+0.015}_{-0.016} $ & $  3.055^{+0.014}_{-0.016} $ & $  3.077\pm 0.032 $ & $  3.063\pm 0.031 $ & $  3.047\pm 0.035 $ & $  3.035\pm 0.033 $ & $  3.059\pm 0.016 $ & $  3.057^{+0.014}_{-0.016} $ & $  3.047\pm 0.017 $ & $  3.047\pm 0.016 $ \\ 
$ m_\mathrm{a}$ [eV]& $ < 0.888 $ & $ < 0.282 $ & $ < 2.27 $ & $ < 1.12 $ & $ < 4.93 $ & $ < 0.987 $ & $ < 0.190 $ & $ < 0.180 $ & $ < 0.388 $ & $ < 0.310 $ \\ 
$ \sum m_\nu$ [eV] & $ < 0.278 $ & $ < 0.156 $ & $ < 2.63 $ & $ < 0.351 $ & $ < 2.18 $ & $ < 0.339 $ & $ < 0.305 $ & $ < 0.163 $ & $ < 0.300 $ & $ < 0.169 $ \\ 
$ H_0  $ & $  66.9^{+1.2}_{-0.73} $ & $  67.90\pm 0.53 $ & $  60^{+7}_{-4} $ & $  67.97\pm 0.77 $ & $  61.1^{+6.3}_{-3.9} $ & $  68.50\pm 0.74 $ & $  67.0^{+1.1}_{-0.73} $ & $  67.81\pm 0.52 $ & $  67.1^{+1.1}_{-0.72} $ & $  67.95\pm 0.54 $ \\ 
$ \sigma_8  $ & $  0.793^{+0.023}_{-0.011} $ & $  0.8052^{+0.0099}_{-0.0075} $ & $  0.656^{+0.11}_{-0.076} $ & $  0.779^{+0.032}_{-0.026} $ & $  0.611^{+0.10}_{-0.081} $ & $  0.755^{+0.031}_{-0.024} $ & $  0.801^{+0.020}_{-0.011} $ & $  0.809^{+0.011}_{-0.0084} $ & $  0.791^{+0.020}_{-0.011} $ & $  0.799^{+0.012}_{-0.0087} $ \\

				\hline \hline
				
		\end{tabular}}
	\end{center}
		\caption{95\%~CL upper bounds on the QCD axion mass and on the sum of neutrino masses and 68\% CL cosmological parameter errors  in the minimal $\Lambda$CDM picture.}
	\label{tab.lCDM+mnu+ma}
\end{table*}

\begin{table*}[htbp!]
	\begin{center}
		\renewcommand{\arraystretch}{2}
		\resizebox{1 \textwidth}{!}{\begin{tabular}{c | c c| c c| c  c|  c c| c c  }

				\hline
				\textbf{Parameter} & \textbf{\nq{Planck}}  &  \textbf{\nq{\\ +BAO}} & \textbf{\nq{ACT}} & \textbf{\nq{\\ +BAO}}  & \textbf{\nq{SPT}} & \textbf{\nq{\\ +BAO}}  & \textbf{ACT+Planck650} & \nq{\\ \textbf{+BAO}} & \textbf{SPT+Planck650} & \nq{\\ \textbf{+ BAO}} \\
				\hline\hline

$ \Omega_\mathrm{b} h^2  $ & $  0.02256\pm 0.00020 $ & $  0.02248\pm 0.00016 $ & $  0.02178\pm 0.00033 $ & $  0.02166\pm 0.00032 $ & $  0.02253\pm 0.00034 $ & $  0.02244\pm 0.00033 $ & $  0.02249^{+0.00017}_{-0.00021} $ & $  0.02242\pm 0.00013 $ & $  0.02261\pm 0.00018 $ & $  0.02248\pm 0.00015 $ \\ 
$ \Omega_\mathrm{c} h^2  $ & $  0.1214\pm 0.0020 $ & $  0.1214^{+0.0016}_{-0.0019} $ & $  0.1182\pm 0.0047 $ & $  0.1237\pm 0.0048 $ & $  0.1163\pm 0.0063 $ & $  0.1232\pm 0.0052 $ & $  0.1203\pm 0.0016 $ & $  0.1209\pm 0.0014 $ & $  0.1210\pm 0.0019 $ & $  0.1211^{+0.0015}_{-0.0018} $ \\ 
$ \tau  $ & $  0.0497\pm 0.0077 $ & $  0.0589\pm 0.0070 $ & $  0.065\pm 0.015 $ & $  0.071\pm 0.014 $ & $  0.065\pm 0.015 $ & $  0.068\pm 0.015 $ & $  0.0475\pm 0.0079 $ & $  0.0566\pm 0.0077 $ & $  0.0488\pm 0.0083 $ & $  0.0552\pm 0.0079 $ \\ 
$ 100\theta_\mathrm{MC}  $ & $  1.04058\pm 0.00037 $ & $  1.04072\pm 0.00034 $ & $  1.04131\pm 0.00077 $ & $  1.04157\pm 0.00075 $ & $  1.03868\pm 0.00085 $ & $  1.03886\pm 0.00081 $ & $  1.04075\pm 0.00033 $ & $  1.04094\pm 0.00029 $ & $  1.04037\pm 0.00035 $ & $  1.04052\pm 0.00032 $ \\ 
$ n_\mathrm{s}  $ & $  0.9727\pm 0.0055 $ & $  0.9697\pm 0.0049 $ & $  1.004\pm 0.020 $ & $  1.014\pm 0.017 $ & $  0.994\pm 0.030 $ & $  0.999\pm 0.022 $ & $  0.9727\pm 0.0058 $ & $  0.9716\pm 0.0043 $ & $  0.9738\pm 0.0056 $ & $  0.9709\pm 0.0049 $ \\ 
$ \log(10^{10} A_\mathrm{s})  $ & $  3.037\pm 0.017 $ & $  3.058\pm 0.014 $ & $  3.046\pm 0.033 $ & $  3.075\pm 0.032 $ & $  3.032\pm 0.039 $ & $  3.049\pm 0.036 $ & $  3.036\pm 0.017 $ & $  3.058\pm 0.016 $ & $  3.034\pm 0.018 $ & $  3.047\pm 0.016 $ \\ 
$ \Omega_k  $ & $  -0.029^{+0.016}_{-0.011} $ & $  0.0010\pm 0.0021 $ & $  -0.169^{+0.070}_{-0.091} $ & $  0.0069^{+0.0057}_{-0.0065} $ & $  -0.057^{+0.075}_{-0.035} $ & $  0.0079^{+0.0060}_{-0.0068} $ & $  -0.091^{+0.047}_{-0.032} $ & $  0.0011\pm 0.0020 $ & $  -0.047^{+0.029}_{-0.017} $ & $  0.0014^{+0.0019}_{-0.0022} $ \\ 
$ m_\mathrm{a}$ [eV]& $ < 1.62 $ & $ < 0.359 $ & $ < 2.18 $ & $ < 1.71 $ & $ < 4.34 $ & $ < 1.71 $ & $ < 1.37 $ & $ < 0.176 $ & $ < 1.96 $ & $ < 0.356 $ \\ 
$ \sum m_\nu$ [eV] & $ < 0.736 $ & $ < 0.183 $ & $  2.0^{+1.4}_{-1.4} $ & $ < 0.604 $ & $ < 2.73 $ & $ < 0.567 $ & $ < 1.23 $ & $ < 0.205 $ & $ < 0.808 $ & $ < 0.224 $ \\ 
$ H_0  $ & $  56.3\pm 4.3 $ & $  68.14\pm 0.73 $ & $  35.8^{+3.3}_{-6.6} $ & $  68.45\pm 0.94 $ & $  50^{+9}_{-10} $ & $  69.07\pm 0.91 $ & $  45.0^{+4.6}_{-6.6} $ & $  68.02\pm 0.69 $ & $  52.7\pm 5.4 $ & $  68.25\pm 0.73 $ \\ 
$ \sigma_8  $ & $  0.701\pm 0.047 $ & $  0.805^{+0.011}_{-0.0085} $ & $  0.459^{+0.037}_{-0.069} $ & $  0.768^{+0.035}_{-0.030} $ & $  0.535^{+0.084}_{-0.12} $ & $  0.745^{+0.034}_{-0.030} $ & $  0.627^{+0.051}_{-0.077} $ & $  0.808^{+0.013}_{-0.0094} $ & $  0.688^{+0.067}_{-0.053} $ & $  0.797^{+0.015}_{-0.0098} $ \\

				\hline \hline
				
		\end{tabular}}
	\end{center}
	\caption{95\%~CL upper bounds on the QCD axion mass and on the sum of neutrino masses and 68\% CL cosmological parameter errors  in the presence of a non-zero curvature component.}
	\label{tab.omkCDM+mnu+ma}
\end{table*}

\begin{table*}[htbp!]
	\begin{center}
		\renewcommand{\arraystretch}{2}
		\resizebox{1 \textwidth}{!}{\begin{tabular}{c | c c| c c| c  c|  c c| c c  }

				\hline
				\textbf{Parameter} & \textbf{\nq{Planck}}  &  \textbf{\nq{\\ +BAO}} & \textbf{\nq{ACT}} & \textbf{\nq{\\ +BAO}}  & \textbf{\nq{SPT}} & \textbf{\nq{\\ +BAO}}  & \textbf{ACT+Planck650} & \nq{\\ \textbf{+BAO}} & \textbf{SPT+Planck650} & \nq{\\ \textbf{+ BAO}} \\
				\hline\hline

$ \Omega_\mathrm{b} h^2  $ & $  0.02245\pm 0.00016 $ & $  0.02252\pm 0.00015 $ & $  0.02137\pm 0.00033 $ & $  0.02152\pm 0.00032 $ & $  0.02249\pm 0.00034 $ & $  0.02251\pm 0.00032 $ & $  0.02234\pm 0.00014 $ & $  0.02239\pm 0.00014 $ & $  0.02247\pm 0.00015 $ & $  0.02252\pm 0.00014 $ \\ 
$ \Omega_\mathrm{c} h^2  $ & $  0.1223^{+0.0016}_{-0.0021} $ & $  0.1208^{+0.0011}_{-0.0014} $ & $  0.1240^{+0.0051}_{-0.0045} $ & $  0.1185^{+0.0024}_{-0.0021} $ & $  0.1186^{+0.0060}_{-0.0047} $ & $  0.1174^{+0.0023}_{-0.0019} $ & $  0.1217^{+0.0014}_{-0.0018} $ & $  0.1205^{+0.0011}_{-0.0014} $ & $  0.1218^{+0.0017}_{-0.0020} $ & $  0.1204^{+0.0012}_{-0.0014} $ \\ 
$ \tau  $ & $  0.0572^{+0.0073}_{-0.0082} $ & $  0.0598\pm 0.0070 $ & $  0.064\pm 0.015 $ & $  0.066\pm 0.015 $ & $  0.066\pm 0.015 $ & $  0.066\pm 0.014 $ & $  0.0543\pm 0.0078 $ & $  0.0551\pm 0.0077 $ & $  0.0549\pm 0.0079 $ & $  0.0556\pm 0.0080 $ \\ 
$ 100\theta_\mathrm{MC}  $ & $  1.04061\pm 0.00035 $ & $  1.04082\pm 0.00031 $ & $  1.04079\pm 0.00075 $ & $  1.04222\pm 0.00064 $ & $  1.03864\pm 0.00084 $ & $  1.03943\pm 0.00065 $ & $  1.04083\pm 0.00032 $ & $  1.04098\pm 0.00028 $ & $  1.04041\pm 0.00033 $ & $  1.04060\pm 0.00028 $ \\ 
$ n_\mathrm{s}  $ & $  0.9677\pm 0.0052 $ & $  0.9700\pm 0.0045 $ & $  0.916^{+0.024}_{-0.027} $ & $  0.986\pm 0.019 $ & $  0.969\pm 0.034 $ & $  1.001^{+0.025}_{-0.029} $ & $  0.9724^{+0.0044}_{-0.0050} $ & $  0.9740^{+0.0040}_{-0.0046} $ & $  0.9703\pm 0.0052 $ & $  0.9722\pm 0.0046 $ \\ 
$ \log(10^{10} A_\mathrm{s})  $ & $  3.057\pm 0.016 $ & $  3.058\pm 0.015 $ & $  3.068\pm 0.031 $ & $  3.055\pm 0.030 $ & $  3.041\pm 0.035 $ & $  3.029\pm 0.034 $ & $  3.053\pm 0.017 $ & $  3.051\pm 0.016 $ & $  3.048\pm 0.017 $ & $  3.046\pm 0.017 $ \\ 
$ \alpha_s  $ & $  -0.0027\pm 0.0071 $ & $  -0.0029\pm 0.0067 $ & $  0.133\pm 0.028 $ & $  0.086\pm 0.029 $ & $  0.048\pm 0.054 $ & $  0.031\pm 0.054 $ & $  0.0083\pm 0.0064 $ & $  0.0080\pm 0.0063 $ & $  0.0004\pm 0.0068 $ & $  0.0003\pm 0.0067 $ \\ 
$ m_\mathrm{a}$ [eV] & $ < 0.661 $ & $ < 0.294 $ & $ < 3.92 $ & $ < 1.53 $ & $ < 4.86 $ & $ < 1.04 $ & $ < 0.580 $ & $ < 0.248 $ & $ < 0.753 $ & $ < 0.308 $ \\ 
$ \sum m_\nu$ [eV] & $ < 0.279 $ & $ < 0.155 $ & $  2.3^{+1.9}_{-1.8} $ & $ < 0.375 $ & $ < 2.06 $ & $ < 0.329 $ & $ < 0.325 $ & $ < 0.172 $ & $ < 0.318 $ & $ < 0.168 $ \\ 
$ H_0  $ & $  67.0^{+1.1}_{-0.76} $ & $  67.93\pm 0.52 $ & $  51.0^{+3.5}_{-5.2} $ & $  68.00^{+0.74}_{-0.66} $ & $  61.0^{+5.8}_{-4.1} $ & $  68.54\pm 0.72 $ & $  66.9^{+1.2}_{-0.72} $ & $  67.79\pm 0.53 $ & $  67.0^{+1.2}_{-0.77} $ & $  67.96\pm 0.54 $ \\ 
$ \sigma_8  $ & $  0.794^{+0.021}_{-0.010} $ & $  0.8057^{+0.0098}_{-0.0072} $ & $  0.489^{+0.045}_{-0.073} $ & $  0.756\pm 0.031 $ & $  0.600^{+0.097}_{-0.083} $ & $  0.753^{+0.031}_{-0.026} $ & $  0.795^{+0.026}_{-0.012} $ & $  0.807^{+0.012}_{-0.0086} $ & $  0.786^{+0.027}_{-0.013} $ & $  0.799^{+0.012}_{-0.0088} $ \\

				\hline \hline
				
		\end{tabular}}
	\end{center}
		\caption{95\%~CL upper bounds on the QCD axion mass and on the sum of neutrino masses and 68\% CL cosmological parameter errors  in the presence of a running of the scalar tilt in the primordial power spectrum.}
	\label{tab.mnu+ma+nrun}
\end{table*}

\begin{table*}[htbp!]
	\begin{center}
		\renewcommand{\arraystretch}{2}
		\resizebox{1 \textwidth}{!}{\begin{tabular}{c | c c| c c| c  c|  c c| c c  }

				\hline
				\textbf{Parameter} & \textbf{\nq{Planck}}  &  \textbf{\nq{\\ +BAO}} & \textbf{\nq{ACT}} & \textbf{\nq{\\ +BAO}}  & \textbf{\nq{SPT}} & \textbf{\nq{\\ +BAO}}  & \textbf{ACT+Planck650} & \nq{\\ \textbf{+BAO}} & \textbf{SPT+Planck650} & \nq{\\ \textbf{+ BAO}} \\
				\hline\hline

$ \Omega_\mathrm{b} h^2  $ & $  0.02247\pm 0.00017 $ & $  0.02248\pm 0.00015 $ & $  0.02166\pm 0.00032 $ & $  0.02169\pm 0.00032 $ & $  0.02249\pm 0.00033 $ & $  0.02252\pm 0.00033 $ & $  0.02241\pm 0.00014 $ & $  0.02242\pm 0.00013 $ & $  0.02249\pm 0.00015 $ & $  0.02251\pm 0.00014 $ \\ 
$ \Omega_\mathrm{c} h^2  $ & $  0.1221^{+0.0018}_{-0.0023} $ & $  0.1215^{+0.0013}_{-0.0018} $ & $  0.1243\pm 0.0045 $ & $  0.1191\pm 0.0026 $ & $  0.1198^{+0.0062}_{-0.0048} $ & $  0.1179\pm 0.0025 $ & $  0.1212\pm 0.0015 $ & $  0.1208\pm 0.0013 $ & $  0.1216^{+0.0016}_{-0.0020} $ & $  0.1210^{+0.0014}_{-0.0017} $ \\ 
$ \tau  $ & $  0.0547\pm 0.0077 $ & $  0.0566\pm 0.0077 $ & $  0.072\pm 0.015 $ & $  0.070\pm 0.015 $ & $  0.067\pm 0.015 $ & $  0.068\pm 0.014 $ & $  0.0556\pm 0.0079 $ & $  0.0559^{+0.0073}_{-0.0081} $ & $  0.0542\pm 0.0075 $ & $  0.0548\pm 0.0078 $ \\ 
$ 100\theta_\mathrm{MC}  $ & $  1.04060\pm 0.00037 $ & $  1.04071\pm 0.00032 $ & $  1.04114\pm 0.00080 $ & $  1.04201\pm 0.00068 $ & $  1.03846\pm 0.00083 $ & $  1.03932\pm 0.00068 $ & $  1.04088\pm 0.00030 $ & $  1.04096\pm 0.00029 $ & $  1.04043\pm 0.00034 $ & $  1.04052\pm 0.00031 $ \\ 
$ n_\mathrm{s}  $ & $  0.9695\pm 0.0050 $ & $  0.9697\pm 0.0046 $ & $  1.003^{+0.022}_{-0.019} $ & $  1.023\pm 0.015 $ & $  0.983^{+0.029}_{-0.026} $ & $  1.013\pm 0.018 $ & $  0.9708\pm 0.0044 $ & $  0.9718\pm 0.0041 $ & $  0.9705\pm 0.0049 $ & $  0.9716\pm 0.0046 $ \\ 
$ \log(10^{10} A_\mathrm{s})  $ & $  3.050\pm 0.016 $ & $  3.053\pm 0.015 $ & $  3.078\pm 0.033 $ & $  3.062\pm 0.031 $ & $  3.049\pm 0.036 $ & $  3.034\pm 0.034 $ & $  3.056\pm 0.016 $ & $  3.056\pm 0.016 $ & $  3.046\pm 0.016 $ & $  3.046\pm 0.016 $ \\ 
$ w_{0}  $ & $  -1.64^{+0.22}_{-0.33} $ & $  -1.052^{+0.061}_{-0.047} $ & $  -1.70\pm 0.63 $ & $  -1.035^{+0.098}_{-0.074} $ & $  -1.55^{+0.80}_{-0.67} $ & $  -1.051^{+0.098}_{-0.075} $ & $  -1.62^{+0.24}_{-0.34} $ & $  -1.039^{+0.057}_{-0.046} $ & $  -1.58^{+0.25}_{-0.36} $ & $  -1.044^{+0.062}_{-0.047} $ \\ 
$ m_\mathrm{a}$ [eV]  & $ < 0.858 $ & $ < 0.466 $ & $ < 2.31 $ & $ < 1.41 $ & $ < 4.96 $ & $ < 1.44 $ & $ < 0.192 $ & $ < 0.181 $ & $ < 0.755 $ & $ < 0.442 $ \\ 
$ \sum m_\nu  $ [eV] & $ < 0.343 $ & $ < 0.221 $ & $ < 2.67 $ & $ < 0.455 $ & $ < 2.30 $ & $ < 0.438 $ & $ < 0.400 $ & $ < 0.232 $ & $ < 0.376 $ & $ < 0.242 $ \\ 
$ H_0  $ & $ > 83.1 $ & $  69.0^{+1.2}_{-1.4} $ & $  72^{+10}_{-20} $ & $  68.5^{+1.6}_{-1.8} $ & $  71^{+10}_{-20} $ & $  69.4^{+1.6}_{-1.9} $ & $ > 82.1 $ & $  68.6^{+1.1}_{-1.3} $ & $ > 80.5 $ & $  68.9^{+1.2}_{-1.4} $ \\ 
$ \sigma_8  $ & $  0.945^{+0.083}_{-0.053} $ & $  0.812\pm 0.014 $ & $  0.73\pm 0.14 $ & $  0.775^{+0.031}_{-0.028} $ & $  0.66^{+0.12}_{-0.14} $ & $  0.751^{+0.031}_{-0.028} $ & $  0.955^{+0.090}_{-0.057} $ & $  0.815\pm 0.015 $ & $  0.926^{+0.092}_{-0.062} $ & $  0.804\pm 0.016 $ \\ 
			
			\hline \hline	

		\end{tabular}}
	\end{center}
		\caption{95\%~CL upper bounds on the QCD axion mass and on the sum of neutrino masses and 68\% CL cosmological parameter errors  varying the dark energy equation of state.}

	\label{tab.w0}
\end{table*}

\begin{table*}[htbp!]
	\begin{center}
		\renewcommand{\arraystretch}{2}
		\resizebox{1 \textwidth}{!}{\begin{tabular}{c | c c| c c| c  c|  c c| c c  }

				\hline
				\textbf{Parameter} & \textbf{\nq{Planck}}  &  \textbf{\nq{\\ +BAO}} & \textbf{\nq{ACT}} & \textbf{\nq{\\ +BAO}}  & \textbf{\nq{SPT}} & \textbf{\nq{\\ +BAO}}  & \textbf{ACT+Planck650} & \nq{\\ \textbf{+BAO}} & \textbf{SPT+Planck650} & \nq{\\ \textbf{+ BAO}} \\
				\hline\hline

$ \Omega_\mathrm{b} h^2  $ & $  0.02248\pm 0.00017 $ & $  0.02245\pm 0.00015 $ & $  0.02166\pm 0.00033 $ & $  0.02168\pm 0.00032 $ & $  0.02249\pm 0.00033 $ & $  0.02249\pm 0.00033 $ & $  0.02241\pm 0.00014 $ & $  0.02239\pm 0.00013 $ & $  0.02252\pm 0.00016 $ & $  0.02247\pm 0.00014 $ \\ 
$ \Omega_\mathrm{c} h^2  $ & $  0.1220^{+0.0018}_{-0.0022} $ & $  0.1223^{+0.0015}_{-0.0020} $ & $  0.1241\pm 0.0045 $ & $  0.1207\pm 0.0029 $ & $  0.1198^{+0.0059}_{-0.0046} $ & $  0.1191\pm 0.0028 $ & $  0.1212^{+0.0014}_{-0.0016} $ & $  0.1214^{+0.0012}_{-0.0014} $ & $  0.1217^{+0.0018}_{-0.0022} $ & $  0.1216^{+0.0015}_{-0.0018} $ \\ 
$ \tau  $ & $  0.0547\pm 0.0076 $ & $  0.0557^{+0.0071}_{-0.0082} $ & $  0.072\pm 0.015 $ & $  0.070\pm 0.015 $ & $  0.067\pm 0.015 $ & $  0.067\pm 0.015 $ & $  0.0554\pm 0.0077 $ & $  0.0556\pm 0.0079 $ & $  0.0542\pm 0.0078 $ & $  0.0545^{+0.0070}_{-0.0078} $ \\ 
$ 100\theta_\mathrm{MC}  $ & $  1.04062\pm 0.00037 $ & $  1.04060\pm 0.00034 $ & $  1.04117\pm 0.00081 $ & $  1.04185\pm 0.00068 $ & $  1.03846\pm 0.00080 $ & $  1.03921\pm 0.00070 $ & $  1.04088\pm 0.00031 $ & $  1.04087\pm 0.00029 $ & $  1.04035\pm 0.00035 $ & $  1.04044\pm 0.00031 $ \\ 
$ n_\mathrm{s}  $ & $  0.9697\pm 0.0051 $ & $  0.9691\pm 0.0047 $ & $  1.004^{+0.022}_{-0.019} $ & $  1.021\pm 0.015 $ & $  0.984^{+0.029}_{-0.025} $ & $  1.010\pm 0.019 $ & $  0.9711\pm 0.0045 $ & $  0.9707\pm 0.0042 $ & $  0.9702^{+0.0058}_{-0.0052} $ & $  0.9705\pm 0.0047 $ \\ 
$ \log(10^{10} A_\mathrm{s})  $ & $  3.050\pm 0.016 $ & $  3.053^{+0.015}_{-0.017} $ & $  3.076\pm 0.032 $ & $  3.066\pm 0.031 $ & $  3.048\pm 0.036 $ & $  3.035\pm 0.034 $ & $  3.056\pm 0.016 $ & $  3.057\pm 0.016 $ & $  3.048\pm 0.017 $ & $  3.047^{+0.015}_{-0.016} $ \\ 
$ w_{0}  $ & $  -1.32^{+0.43}_{-0.55} $ & $  -0.70\pm 0.21 $ & $  -1.46^{+0.72}_{-1.1} $ & $  -0.80\pm 0.25 $ & $  -1.30^{+0.83}_{-1.1} $ & $  -0.88\pm 0.24 $ & $  -1.30^{+0.44}_{-0.58} $ & $  -0.70\pm 0.22 $ & $  -1.10^{+0.47}_{-0.70} $ & $  -0.74\pm 0.22 $ \\ 
$ w_{a}  $ & $ < -0.693 $ & $  -1.14^{+0.79}_{-0.65} $ & $ < -0.325 $ & $  -0.96^{+1.0}_{-0.85} $ & $ < -0.284 $ & $  -0.70^{+1.0}_{-0.74} $ & $ < -0.671 $ & $  -1.10^{+0.80}_{-0.64} $ & $ < -1.19 $ & $  -0.99^{+0.80}_{-0.63} $ \\ 
$ m_{\rm a}$ [eV] & $ < 0.972 $ & $ < 0.716 $ & $ < 2.36 $ & $ < 1.66 $ & $ < 4.85 $ & $ < 1.59 $ & $ < 0.222 $ & $ < 0.204 $ & $ < 2.60 $ & $ < 0.544 $ \\ 
$ \sum m_\nu$ [eV]& $ < 0.337 $ & $ < 0.291 $ & $ < 2.66 $ & $ < 0.549 $ & $ < 2.26 $ & $ < 0.505 $ & $ < 0.378 $ & $ < 0.326 $ & $ < 0.420 $ & $ < 0.305 $ \\ 
$ H_0  $ & $ > 80.4 $ & $  66.7^{+1.7}_{-2.0} $ & $  71^{+10}_{-20} $ & $  67.3^{+1.9}_{-2.3} $ & $  70^{+10}_{-20} $ & $  68.4^{+2.1}_{-2.3} $ & $ > 79.6 $ & $  66.2\pm 1.9 $ & $ > 75.3 $ & $  66.8\pm 1.9 $ \\ 
$ \sigma_8  $ & $  0.929^{+0.10}_{-0.060} $ & $  0.791\pm 0.019 $ & $  0.72\pm 0.14 $ & $  0.766^{+0.034}_{-0.030} $ & $  0.65^{+0.12}_{-0.14} $ & $  0.745\pm 0.031 $ & $  0.941^{+0.11}_{-0.063} $ & $  0.796\pm 0.020 $ & $  0.880^{+0.14}_{-0.070} $ & $  0.789\pm 0.020 $ \\ 
				\hline \hline
				
		\end{tabular}}
	\end{center}
	\caption{95\%~CL upper bounds on the QCD axion mass and on the sum of neutrino masses and 68\% CL cosmological parameter errors  varying the dark energy equation of state using the CPL two-parameter parameterization, see Eq.~(\ref{eq:cpl}).}
	\label{tab.w0wa}
	\end{table*}

\section{Bayesian evidences and Bayes factors}
\label{sec:appendixBEBF}
\begin{figure*}
\centering
\includegraphics[width=0.85\textwidth]{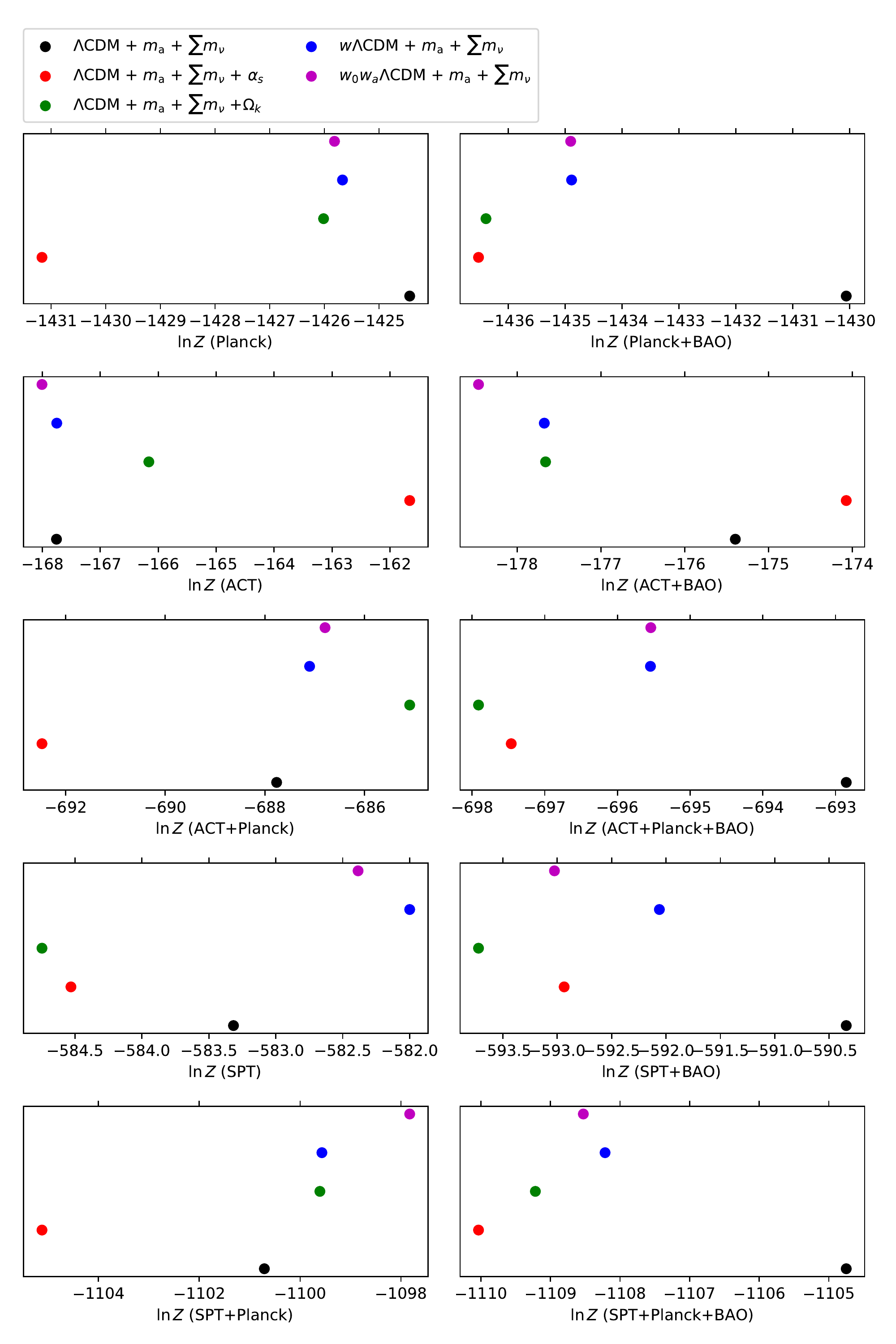}
\caption{Bayesian evidences for each model and dataset combination.
\label{fig:BEs}}
\end{figure*}

\begin{figure*}
\centering
\includegraphics[width=0.7\textwidth]{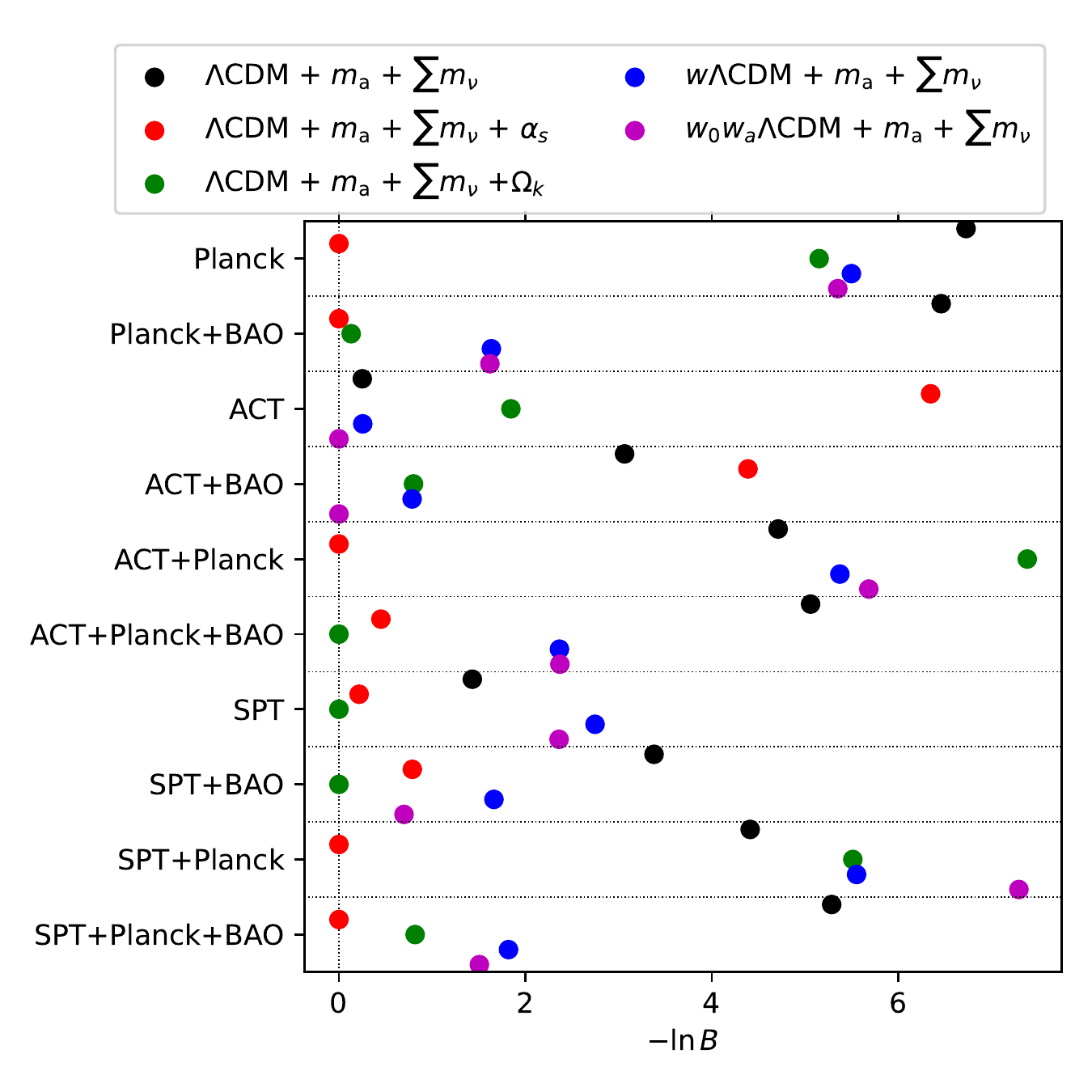}
\caption{Bayes factors, normalized to the best model for each dataset.
\label{fig:BFs}}
\end{figure*}

In this appendix we provide a figure (\autoref{fig:BEs}) reporting the Bayesian evidences for each model and within each dataset combination, and a figure (\autoref{fig:BFs}) representing the Bayes factors listed in table~\ref{tab:BFs}.

\clearpage 
\bibliography{biblio}	
\end{document}